\documentclass{article}

\usepackage[a4paper]{geometry}
\usepackage[english]{babel}
\usepackage{natbib}
\usepackage{url}
\usepackage{booktabs}
\usepackage{setspace}
\usepackage{pgffor}
\usepackage{nicefrac}
\usepackage{multirow}
   
\usepackage{psfrag}
\usepackage{graphicx}
\usepackage{subfigure}
\usepackage{psfrag}
\usepackage{latexsym}
\usepackage{float}
\usepackage{enumerate}
\usepackage{bbold}
\usepackage{amsmath}
\usepackage{pxfonts}

\usepackage{rotating}
\usepackage{xcolor}

\newcommand{\optional}[1]{\textcolor{gray}{#1}}
\renewcommand{\optional}[1]{{#1}}

\newcommand{\R}{\mathbb{R}}

\newcommand{\T}{\top}
\renewcommand{\vec}[1]{\big(#1\big)^\T}

\newcommand{\zero}{\boldsymbol{0}}

\newcommand{\bA}{\mathbf{A}}
\newcommand{\ba}{\mathbf{a}}
\newcommand{\bB}{\mathbf{B}}
\newcommand{\bC}{\mathbf{C}}
\newcommand{\bD}{\mathbf{D}}

\newcommand{\bH}{\mathbf{H}}
\newcommand{\bh}{\mathbf{h}}
\newcommand{\bL}{\mathbf{L}}
\newcommand{\bP}{\mathbf{P}}
\newcommand{\bQ}{\mathbf{Q}}
\newcommand{\bR}{\mathbf{R}}
\newcommand{\br}{\mathbf{r}}
\newcommand{\bb}{\mathbf{b}}
\newcommand{\bZ}{\mathbf{Z}}

\newcommand{\bW}{\mathbf{W}}

\newcommand{\bt}{\mathbf{t}}

\newcommand{\bx}{\mathbf{x}}
\newcommand{\bX}{\mathbf{X}}

\newcommand{\bI}{\mathbf{I}}

\newcommand{\btheta}{\boldsymbol{\theta}}
\newcommand{\bvartheta}{\boldsymbol{\vartheta}}

\newcommand{\bOmega}{\boldsymbol{\Omega}}

\newcommand{\bzeta}{\boldsymbol{\zeta}}

\newcommand{\E}{\mathbb{E}}
\newcommand{\Cov}{\mathbf{Cov}}
\newcommand{\Var}{\mathbb{V}\mathrm{ar}}
\newcommand{\argmin}[1][ ]{\underset{#1}{\operatorname{argmin}\;}}
\newcommand{\KL}{\operatorname{KLD}}
\newcommand{\tr}{\operatorname{tr}}
\newcommand{\extinctprop}{p}
\newcommand{\steplen}{\nu}
\newcommand{\intervalT}{\mathcal{T}}
\newcommand{\intervalS}{\mathcal{S}}

\newcommand{\sd}[1]{\operatorname{sd}\! \left( #1 \right)}

\newcommand{\jq}{_{j}^{(q)}}
\newcommand{\jXq}{_{Xj}^{(q)}}
\newcommand{\jYq}{_{Yj}^{(q)}}
\newcommand{\df}{d\! f}
\begin{document}

\title
{Boosting Functional Response Models for Location, Scale and Shape with an Application to Bacterial Competition}
\author{Almond St\"ocker$^1$ \and Sarah Brockhaus$^1$ \and Sophia Schaffer$^2$ \and 
	Benedikt von Bronk$^2$ \and Madeleine Opitz$^2$ \and Sonja Greven$^1$}
\date{
	$^1$Department of Statistics, LMU Munich, Germany\\%
	$^2$Faculty of Physics, LMU Munich, Germany\\%
}

\index{St\"ocker, A.} 

\maketitle

\begin{abstract} 
	We extend Generalized Additive Models for Location, Scale, and Shape (GAMLSS) to regression with functional response. 
	This allows us to simultaneously model point-wise mean curves, variances and other distributional parameters of the response in dependence of various scalar and functional covariate effects. 
	In addition, the scope of distributions is extended beyond exponential families. 
	The model is fitted via gradient boosting, which offers inherent model selection and is shown to be suitable for both complex model structures and highly auto-correlated response curves. 
	This enables us to analyze bacterial growth in \textit{Escherichia coli} in a complex interaction scenario, fruitfully extending usual growth models.\\
	
\end{abstract}


	\section{Introduction}
\label{chap_intro}

In functional data analysis \citep{RamsaySilverman2005}, functional response regression aims at estimating covariate effects on response curves \citep{Morris2015, GrevenScheipl2017}. The response curves might, for instance, be given by annual temperature curves, growth curves or spectroscopy data, to name a few popular examples.  
We propose a flexible approach to regression with functional response allowing for simultaneously modeling multiple distributional characteristics of response curves. It therefore generalizes usual functional mean regression models and poses another step in recent efforts to extend flexible regression models to a unified general functional regression framework \citep{GrevenScheipl2017}.

We apply this framework to analyze bacterial interaction: as bacterial resistances increase, producing effective antibiotics gets harder and harder. Understanding bacterial interactions might help finding alternatives. In particular, we analyze growth curves of two competing \textit{Escherichia coli} bacteria strains \citep{vonBronk2016} -- a toxin producing `C-strain' and a toxin sensitive `S-strain' -- to obtain insights into the underlying growth affecting bacterial interaction. Both strains are exposed to different external stress levels by adding the antibiotic Mitomycin C to the nutrient supply in the petri dish the strains are growing on. Our aim is to model the S-strain growth behavior in dependence on the toxin emitting C-strain and the stress level, and allow this dependence to affect both mean and variability of growth as well as the extinction probability. This requires a functional response regression model for several parameters of the response distribution with linear and smooth effects of functional and scalar covariates.

There are various approaches applied to modeling bacterial growth curves in the literature. 
Gompertz and Baranyi-Roberts models are two common parametric approaches to modeling growth curves (see, e.g., \citet{LopezFrance2004} and \citet{Perni2005}). \citet{Weber2014} implement a model particularly for analyzing bacterial interaction. The models are usually fitted using least squares methods, which corresponds to assuming a Gaussian distribution of bacterial propagation. 
This is problematic as response values are naturally positive and very small in the beginning, starting from single cell level. Thus, also assuming a constant variance over the whole time span seems not appropriate. Moreover, they do not offer the opportunity to include covariate effects for modeling, e.g., the impact of the applied external stress. In some cases, a possible two-step approach would be to fit individual parametric models to each curve and then perform regression for the parameters. However, with time-varying covariates as in our case, this is not an option. Thus, these models are not applicable here. 
In addition, they are often highly non-linear, which may introduce problems in parameter estimation. \citet{GasserMuellerEtAl1984} discuss this point and some further advantages of nonparametric growth curve regression over parametric models. They propose a kernel method for this purpose, which does, however, not include covariates and lacks the flexibility needed for our purposes. Still, nonparametric functional regression models present a natural choice from a statistical perspective, also because they can approximate the above parametric growth models very well when these are appropriate (see Online Supplement D.2). 

The problem of non-Gaussian functional response also appears in many other applications and is, accordingly, addressed in several publications. Following early works on non-Gaussian functional data by \cite{Hall2008} and \cite{Linde2009}, authors like \cite{Goldsmith2015}, \cite{Wang2014}, and \cite{Scheipl2016} proposed Generalized Linear Mixed Model (GLMM)-type regression models, which are suitable for, e.g., positive, discrete or integer valued response functions. The linear predictor for the mean function is typically composed of a covariate effect term and a latent random Gaussian error process accounting for auto-correlation, and combined with a link function.
\cite{Li2014} jointly model continuous and binary valued functional responses in a similar fashion, but without considering covariates. 
Other ideas to account for the auto-correlation of functional response curves include robust covariance estimation for valid inference after estimation under a working independence assumption \citep{Gertheiss2015} and an over-all regularization by early-stopping a gradient boosting fitting procedure guided by curve-wise re-sampling methods as applied by \citep{BrockhausGreven2015, BrockhausGreven2016a} besides using curve-specific smooth errors. 
Moreover, this latter approach also offers quantile regression for functional data. 
The above models present important steps towards meeting the challenges in the bacterial interaction scenario. In particular, our approach is a direct generalization of the framework of \cite{BrockhausGreven2015}. However, the previous methods are restricted to one predictor such that none of them allows for simultaneously modeling also the response variance or other distributional parameters in a similar fashion as the mean.
\citet{Staicu2012} propose a method for estimating mean, variance and other shape parameter functions nonparametrically and also for non-Gaussian point-wise distributions, while modeling auto-correlation via copulas. However, they do not allow for including covariate effects.
Only the framework of \cite{Scheipl2016} now allows to perform simultaneous mean and variance regression in the Gaussian case \citep{GrevenScheipl2017}. We will use their model for a simulation based comparison in this special case - also because it is the only one offering the required range of smooth/linear effects of scalar and functional covariates, which are implemented in the R package \texttt{refund} \citep{refund}.
However, in the bacterial interaction data the curves are positive valued and non-Gaussian. In addition, besides modeling mean growth curves and their variance in dependence of covariates, we also have to model the positive probability for the response value being exactly zero, if the respective bacterial strain completely vanishes.

We address these challenges more generally by introducing Generalized Additive Models for Location, Scale and Shape (GAMLSS) for functional responses, which are also suitable for a wide range of other model scenarios. 
For the case of scalar response regression, GAMLSS were introduced by \cite{RigbyStasinopolous2005} extending usual Generalized Additive Models \citep[GAMs; ][]{HastieTibshirani1990} to multiple distributional parameters. Each parameter of the assumed response distribution is modeled with a separate predictor depending on covariates, allowing, e.g., for covariate effects on mean and variance. Hence, doubtful assumptions of homoscedasticity can be overcome. In addition to this extension, the range of applicable distributions of GAMs is also extended to non-exponential family distributions in the GAMLSS framework. This is also important in our case, as it gives us the opportunity to specify a zero-adjusted distribution to account for the positive probability of zero bacterial area.
Combining a scalar GAMLSS framework developed by \citet{MayrSchmid2012} and \citet{Thomas2016} and the flexible functional regression framework of \citet{BrockhausGreven2016a}, \citet{BrockhausGreven2016b} discussed GAMLSS scalar-on-function regression based on the flexible gradient boosting regression framework introduced by \cite{BuehlmannHothorn2007}. We further extend this to functional GAMLSS for function-on-scalar and function-on-function regression, such that our framework now offers the full flexibility of scalar GAMLSS also for functional responses and covariates. In particular, it meets all the challenges arising from the present analysis of bacterial interaction.
The approach is implemented in the R \citep{R} add-on package \texttt{FDboost} \citep{FDboost}.

The remainder of the paper is structured as follows: In Section \ref{chap_Model} we formulate the general model and describe the fitting algorithm. In Section \ref{chap_Application} we apply the proposed model to analyzing \textit{E. coli} bacteria growth. Section \ref{chap_Simulation} provides the results of two simulation studies for Gaussian response curves as well as for the used bacterial growth model. Section \ref{chap_Discussion} concludes with a discussion. Further details concerning the model, application and simulation studies are provided as Online Supplement, as well as the code for the simulations and fitting of the model with the R-package \texttt{FDboost}. 
	
\section{Model formulation}
\label{chap_Model}

Consider a data scenario with $N$ observations of a functional response $Y$ and respective covariates $\bX$. $Y$ is a stochastic process, such that its realized trajectories $y_i : \intervalT \rightarrow \R,$ $t \mapsto y_i(t)$ for $i = 1,...,N$ represent the response curves over an index set $\intervalT$.  
For notational simplicity, we assume that response curves are observed on a common grid $\intervalT_0 \subset \intervalT$, where $\intervalT = [0, t_{max}]$ is a real interval starting at zero and $\intervalT_0$ a finite discrete set of evaluation points. However, the curves could be measured on different grids as well. As this is the case in many applications, the variable $t$ is referred to as time variable. Scalar response is contained as special case where $\intervalT$ is a one point set.
Let $\bx_i = (x_{i,1},..., x_{i,p})^\T$ denote the $i$-th observed covariate vector, i.e. realization of $\bX$, which can contain scalar and functional covariates. A functional covariate may have a different domain $\intervalS$ from the response and is denoted as $x_{i,j}: \intervalS \rightarrow \R$, $s \mapsto x_{i,j}(s)$. We suppress potential dependence of $\intervalS$ on $j$ in our notation.

We assume that for all $t\in \intervalT$ the point-wise distribution of the response $\mathcal{F}_{Y(t)|\bX}$ is known up to the distribution parameters $\bvartheta(t)=\left(\vartheta^{(1)}(t), ..., \vartheta^{(Q)}(t)\right)^\T$. For instance, for a Gaussian process, the parameters might represent the conditional mean and variance over time, i.e., $\vartheta^{(1)}(t) = \E(Y(t)|\bX = \bx)$ and $\vartheta^{(2)}(t) = \Var(Y(t)|\bX = \bx)$, suppressing the dependence on $\bx$ in the notation. For each parameter an additive regression model is assumed. The model is specified by
\begin{equation*}
g^{(q)}(\vartheta^{(q)}) = h^{(q)}(\bx) = \sum_{j = 1}^{J^{(q)}} h\jq(\bx) \hspace{1cm},\ q = 1,...,Q ,
\end{equation*}
where $g^{(q)}$ is a monotonic link function for the $q$-th distribution parameter. This model structure corresponds to the GAMLSS introduced by \citet{RigbyStasinopolous2005}. However, covariates and response may now be functions. Correspondingly, $\vartheta^{(q)}:=\vartheta^{(q)}(.)$ and the predictor $h^{(q)}(\bx) := h^{(q)}(\bx, .)$ are now functions over the domain $\intervalT$ of the response, for $q=1,...,Q$. For $Q=1$ parameter corresponding to the mean, the model reduces to the functional additive regression model of \citet{Scheipl2015, Scheipl2016, BrockhausGreven2015}.

Both covariate and time dependency are modeled within the additive predictor via the effect functions $h^{(q)}_j(\bx,t)$. 
The predictor is typically composed of a functional intercept $h_1^{(q)}(\bx,t) = \beta_0(t)$ and linear or smooth covariate effects $h\jq(\bx,t)$, of which each depends on one or more covariates. The construction of the effects follows a modular principle, which allows for flexible specification of effect types. Table~\ref{effects} gives an overview of different effect types available.

Apart from the Gaussian a variety of other distributions can be specified for the response. 
In principle, $\mathcal{F}_{Y(t)|\bX}$ can be any distribution for which both the likelihood and its derivatives are computable. The derivatives with respect to the parameters are required for the model estimation via gradient boosting. For functional response, usually only continuous distributions are under consideration as response curves are typically assumed to be continuous. However, this does not necessarily have to be the case (see e.g., \citet{Scheipl2016}). \citet{RigbyStasinopolous2005} offer a comprehensive list of distributions, implemented in R. As we built on the approach of \cite{MayrSchmid2012} for scalar GAMLSS, all of these are available for the present boosting approach and further distributions can be manually specified. 

\begin{table}
	\caption{Overview of possible effect types \citep[adapted from][]{BrockhausGreven2015} 
	\label{effects}}
	\small
	\centering
	\begin{tabular} {l c r}
		\toprule
		Covariate(s) & Type of effect & $h\jq$ \\
		\midrule
		(none) & Smooth intercept & $\beta_0(t)$ \\[.3cm]
		Scalar covariate $z$ & Linear effect & $z\, \beta(t)$ \\
		&  Smooth effect & $f(z, t)$ \\[.3cm]
		Two scalars $z_1, z_2$ & Linear interaction & $z_1\, z_2\, \beta(t)$ \\
		& Functional varying coefficient & $z_1\, f(z_2, t)$ \\
		& Smooth interaction & $f(z_1, z_2, t)$ \\
		\hline
		Grouping variable $g$ & Group-specific intercept & $\beta_g(t)$ \\[.3cm]
		Group. variable $g$, scalar $z$ & Group-specific linear effect & $z\, \beta_g(t)$ \\
		& Group-specific smooth effect & $f_g(z, t)$ \\[.3cm]
		Group. variables $g_1, g_2$ & Group-interaction & $\beta_{g_1, g_2} (t)$ \\
		\hline
		Functional covariate $x(s)$ & Functional linear effect & $\int x(s)\, \beta(s,t) \, ds$ \\[.3cm]
		Functional cov. $x(s)$, scalar $z$ & Linear interaction & $z \int x(s)\, \beta(s,t) \, ds$ \\
		& Smooth interaction & $\int x(s)\, \beta(z,s,t) \, ds$ \\[.3cm]
		Functional cov. $x(s)$ over $\intervalT$ & Concurrent effect & $x(t) \beta(t)$ \\
		 & Historical effect & $\int_{0}^{t} x(s)\, \beta(s,t) \, ds$ \\
		 & \shortstack{Effect with $t$-specific\\integration limits} & $\int_{l(t)}^{u(t)} x(s)\, \beta(s,t) \, ds$ \\
		\bottomrule
	\end{tabular}
\end{table}

\subsection{Construction of effect functions $\mathbf h\jq$}
\label{section_effects}
As both, covariate and time dependency of the functional response are specified by the effect functions $h\jq$, they play a key role in the framework.
We briefly illustrate their modular structure and refer to \cite{FDboost} and \cite{GrevenScheipl2017} for further examples and details, as their construction is not new to the functional GAMLSS.
The novelty is, however, that we may now use them in multiple predictors for multiple parameters.\\ 
For each effect type, $h\jq$ is represented by a linear combination of specified basis functions, such that the predictor is linear in its coefficients. Multivariate basis functions are constructed as tensor products of univariate bases providing flexible modular means of specification \citep[cf.][]{Scheipl2015}, giving the basis representation
\begin{equation}
\label{effectformula}
h\jq(\bx, t) = \left(\bb\jXq(\bx, t) \otimes \bb\jYq(t)\right)^\T \btheta^{(q)}_j \hspace{1cm} , t\in \intervalT \, . 
\end{equation}
A vector $\bb\jYq$ of $K\jYq$ basis functions for the time variable is combined with a vector $\bb\jXq$ of $K\jXq$ basis functions for the covariate effects. The basis $\bb\jXq(\bx, t)$ might be time dependent, e.g., for a functional historical effect. However, for many effect types, it only depends on covariates, such that we can write $\bb\jXq(\bx)$. Applying the Kronecker product $\otimes$, a new basis is obtained. \optional{Its elements correspond to the pairwise products of elements in $\bb\jYq$ and $\bb\jXq$.} For details see Online Supplement A.1.
The coefficient vector $\btheta\jq \in \R^{K\jYq K\jXq}$ specifies 
the concrete form of the effect. Fitting the model corresponds to estimating $\btheta\jq$ for all effect functions.

A typical choice for $\bb\jYq(t)$ is a spline basis. Then, in case of time-independent covariate basis functions $\bb\jXq$, $h\jq(\bx_0, t)$ describes a spline curve for a fixed value $\bx = \bx_0$ of the covariate. Usually, quadratic penalty terms are employed in order to control smoothness of the effect functions (see Section \ref{section_fit} Model fit). 
\optional{A typical effect function $h\jq(\bx, t)$ depends on a single covariate. For example, for a linear effect $z\beta(t)$ of a scalar covariate $z$, this yields $\bb\jXq(\bx,t) = \bb\jXq(z) = z$. In order to obtain a smooth covariate effect $f(z,t)$, a spline basis can be chosen for $\bb\jXq(z)$ just like for the time curve, yielding a tensor product spline basis in \eqref{effectformula}.} For a functional covariate $x: \intervalT \mapsto \R$ a historical effect of the form $\int_{0}^{t} x(s)\, \beta(s,t) \, ds$ can be constructed using a basis of linear functionals induced by a tensor product spline basis for $\beta$: the $k$-th basis element is specified by $\bb_{j,k}^{(q)}(x, t) = \int_0^{t} x(s) \varphi_k(s) \, ds$, where $\varphi_k(s),\; k = 1,...,K\jXq$, are the elements of a spline basis, or a common approximation using numerical integration over the observation grid of $x$ in $\intervalT$. 
	


\subsection{Model fit}
\label{section_fit}

The model coefficients are estimated using gradient boosting, an ensemble method with origin in machine learning (\citet{BuehlmannHothorn2007}, \citet{Mayr2014a}). \cite{MayrSchmid2012} developed a model based gradient boosting approach for fitting scalar GAMLSS, which was further refined by \cite{Thomas2016}. \cite{BrockhausGreven2016b} extend the approach to functional covariates. The following presents a full functional GAMLSS also allowing for functional responses. 

Component-wise gradient boosting is a gradient descend method for model fitting, where the model is iteratively updated. In each iteration, the algorithm aims at minimizing a loss function following the direction of its steepest descent. Instead of updating the full additive predictor at once, the individual effect functions $h\jq$ are separately fit to the negative gradient in a component-wise approach. These individual effect models are called \textit{base-learners}, as they present simple base models that jointly form the model predictor. In each iteration, only the effect function providing the largest loss reduction is updated with a step length $\nu$ in the direction of its fit. The component-wise and stepwise procedure yields automatic model selection and allows for fitting models with more parameters than observations. 

Let $f\left(y(t)\big|\bvartheta(t)\right) = f_h\left(y\left(t\right)\, \big|\, \bh\left(\bx,t\right) \right)$ with $\bh = (h^{(1)}, \dots, h^{(Q)})^\T$ denote the conditional probability density function (PDF) of the response at $t \in \intervalT$ for a given parameter setting. We define the point-wise loss function to be the negative log-likelihood
\begin{equation*}
\varrho\left(y\left(t\right), \bh\left(\bx,t\right)\right) = - \log f_h\left(y\left(t\right)\, \big|\, \bh\left(\bx,t\right) \right)\ .
\end{equation*}

\noindent The functional GAMLSS loss function is then obtained as  
\begin{equation*}
\label{ell}
\ell(y, \bh(\bx)) = \int_{\intervalT} \varrho\left( y\right(t\left), \bh\right(\bx,t\left) \right) dt \, ,
\end{equation*} 
the integral over the point-wise loss functions over $\intervalT$. Therefore, we assume that $f_h$ and $\bh$ are chosen such that the integral exists, which is no restriction in practice. 

\noindent The aim of gradient boosting is to find the predictor
\begin{equation}
\label{aim}
	\bh_{optimal} = \argmin[\bh] \E\left[\, \ell(Y, \bh(\bX)) \,\right]  = \argmin[\bh] \int_{\intervalT} \E\left[\, \varrho\left( Y\right(t\left), \bh\right(\bX,t\left) \right) \,\right] dt
\end{equation}
minimizing the expected loss.

Based on data $(y_i, \bx_i)_{i=1,...,N}$, this is estimated by optimizing the empirical mean loss. Hence, the estimated predictor vector $\hat{\bh} = (\hat{h}^{(q)}, ..., \hat{h}^{(q)})^\T$ is given by 
\begin{equation}
\label{minimization}
\hat{\bh} \approx \argmin[\bh] \frac{1}{N} \sum_{i=1}^{N} \hat{\ell}(y_i, \bh(\bx_i)) \, ,
\end{equation}
where $\hat{\ell}(y_i, \bh(\bx_i)) =  \sum_{t\in\intervalT_0} \varrho\left( y_i\right(t\left), \bh\right(\bx_i,t\left) \right)$ is an approximation of the loss.
However, to avoid over-fitting, the optimization is generally not run until convergence. Instead, a re-sampling strategy is employed to find an optimal stopping iteration.

The minimization in \eqref{aim} can be seen to minimize the Kullback-Leibler divergence (KLD) of the model density $f_h$ to the true underlying density.
\cite{HastieTibshirani1990} formulate a similar regression aim for GAMs. However, in the functional case, we consider the point-wise KLD integrated over the domain $\intervalT$.

The base-learners \optional{fitted in each boosting iteration} correspond to the effects $h\jq(\bx, t)$ with $j = 1,...,J^{(q)}$ and $q = 1,...,Q$. For any given loss function, they represent single regression models, which are fitted to the gradient of the loss function via penalized least-squares. The coefficients $\btheta\jq$ of the respective $h\jq$, as defined in equation \eqref{effectformula}, are subject to a quadratic penalty of the form $(\btheta\jq)^\T\bP\jq\btheta\jq$, where $\bP\jq$ is a penalty matrix. As described for bivariate smooth terms, e.g., in \citet{Wood2006} or \cite{BrockhausGreven2015}, the penalty matrix is constructed as
\begin{equation*}
	\bP\jq = \lambda\jXq \left(\bP\jXq \otimes \bI_{K\jYq}\right) + \lambda\jYq \left(\bI_{K\jXq} \otimes \bP\jYq \right)
\end{equation*}
with smoothing parameters $\lambda\jYq, \lambda\jXq \geq 0$ and penalty matrices $\bP\jYq \in \R^{K\jYq \times K\jYq}$ and $\bP\jXq \in \R^{ K\jXq \times K\jXq}$ for the time basis $\bb\jYq(t)$ and covariate basis $\bb\jXq(\bx, t)$, respectively.
For instance, a common choice for B-spline bases is a a first or second order difference penalty matrix yielding P-Splines (compare \cite{EilersMarx2010}). Base learners for group effects might be regularized with a ridge penalty. 
If no penalization should be applied for either the response or the covariates, this can also be obtained by setting $\lambda\jYq = 0$ or $\lambda\jXq = 0$, respectively. 
\cite{Thomas2016} compare different gradient boosting methods for GAMLSS, which can all be analogously generalized to functional response. While the 'cyclic' method and a 'non-cyclic' method are available in the \texttt{R} package \texttt{gamboostLSS}, only the algorithm of the 'non-cyclic' method is described here in detail. Comparing it to the 'cyclic' method, it performed better in simulations (Online Supplement Table 2) 
and provides the advantage of unified model selection across parameters $\vartheta^{(q)}$, $q=1,\dots, Q$. 

\vspace*{.3cm}
{
	\small
\newcommand{\best}[1]{{\tilde{#1}}}
\newcommand{\bestq}[1]{{_q\kern-.5em^\star\! #1}}

\textit{ Algorithm: gradient boosting for functional GAMLSS } \\ 
\hrule
\begin{enumerate}
	\item \textbf{To set up the model specify}
	\begin{enumerate}[i)]
		\item \textbf{a functional loss function $\ell$} with point-wise loss $\varrho$ corresponding to the assumed response distribution with $Q$ distribution parameters
		\item \textbf{the base-learners} by choosing the desired bases for the effects $h\jq(\bx, t) = (\bb\jXq(\bx,t) \otimes \bb\jYq(t))^\T \btheta^{(q)}_j$, penalty matrices $\bP_{j}^{(q)}$ for all $j=1,...,J^{(q)}$ and $q=1,...,Q$ and their respective smoothing parameters.
		\item \textbf{gradient boosting hyper-parameters}: the step-lengths $\steplen^{(q)} \in\, ]0,1]$ for $q = 1,...,Q$ and the maximum number of iterations $m_{stop}$. 
	\end{enumerate}
	Initialize the coefficients $\btheta^{(q)[0]}_j$ for the initial predictor $\bh^{[0]}(\bx_i, t)$, e.g. to $\boldsymbol{0}$, and set $m=0$.
	
	\item \textbf{For $m=0,...,m_{stop}-1$ iterate:}
	\begin{enumerate}[i)]
		\item \textbf{Find best update for each distribution parameter.} \\ \textbf{For $q=1,...,Q$ do:}
		\begin{itemize}
			\item \textbf{Evaluate negative partial gradients }
			for $i=1,...,N$ at the current predictor $\bh^{[m]}$
			\begin{equation*}
			u^{(q)}_i(t):= -\frac{\partial \varrho}{\partial h^{(q)}} \left( y_i (t) , \bh \right) \ \biggr|_{\bh = \bh^{[m]}(\bx_i, t)}
			\end{equation*} 
			
			\item \textbf{Fit base-learners to the gradients},
			i.e., for $j=1,...,J^{(q)}$ find $\best{\btheta}_j^{(q)}$ with
			\begin{align*}
			\best{\btheta}_j^{(q)} :=  \argmin[\btheta\jq] & \Big\{ \sum_{i=1}^N \sum_{t\in \intervalT_0} \left(\; u^{(q)}_i(t) \ \boldsymbol{-} \ \left(\bb\jXq(\bx_i, t) \otimes \bb\jYq(t)\right)^\T \btheta\jq \;\right)^2 \\  &+ (\btheta\jq)^\T\bP_{j}^{(q)}\btheta\jq \Big\}
			\end{align*}
			
			\item \textbf{Determine the best-fitting base-learner} with index $\best{\jmath}$ following the least squares criterion 
			\begin{equation*}
				\best{\jmath} := \argmin[j] \sum_{i=1}^N \sum_{t\in \intervalT_0} \left(\; u^{(q)}_i(t) \ \boldsymbol{-} \ (\bb\jXq(\bx_i, t) \otimes \bb\jYq(t))^\T \best{\btheta}_j^{(q)} \;\right)^2
			\end{equation*}

			\item \textbf{Determine updated predictor candidate, }
			i.e., determine $\bestq{\bh}$ where only the coefficients of the best-fitting base-learner are updated, such that the coefficients are given by
			\begin{equation*}
			\bestq{\btheta}_k^{(p)} = 
			\begin{cases}
			\ \btheta_{k}^{(p)[m]} + \steplen^{(p)}\, \best{\btheta}_{k}^{(p)} & \text{for }p=q, k= \best{\jmath}, \\
			\ \btheta_k^{(p)[m]} & \text{else}
			\end{cases}
			\end{equation*}
		\end{itemize}
		\item[\null] \textbf{end for.}	
			
		\item \textbf{Select best update across the distributional parameters} and update the linear predictor accordingly 
		\begin{equation*}
		\bh^{[m+1]} = \argmin[{_q\kern-.4em^\star\!\bh}] \sum_{i=1}^N \hat{\ell} \left(y_i, \bestq{\bh}(\bx_i)\right)
		\end{equation*}
	\end{enumerate}	
	\textbf{\,\, end for.}	
	
\end{enumerate}
\hrule
}
\vspace{.3cm}
The smoothing parameters for the penalty matrices $\bP_{jY}^{(q)}$ can be chosen indirectly specifying the base-learner degrees of freedom, as described by \citet{HofnerSchmid2011}. They are typically specified such that equal degrees of freedom for all base-learners are attained to ensure a fair base-learner selection. Note that these degrees of freedom only specify the flexibility of each base-learner for one iteration, while the final effective degrees of freedom can be higher due to repeated selection of the same base-learner. $\steplen = 0.1$ is a popular choice for the step-length \citep{BuehlmannHothorn2007}. It should be chosen small enough to prevent overshooting. Yet, too small values greatly increase computation time.  
The optimal stopping iteration $m_{stop}$, with respect to equation \eqref{aim}, is the main tuning parameter. It can be estimated using, e.g., curve-wise cross-validation or bootstrapping. As $\bh^{[m_{stop}]}$ involves computation of all earlier predictors, this can be done very efficiently (and in parallel over cross-validation folds). Early stopping induces shrinkage of effect functions and provides automatic model selection: effect functions $h\jq$ which were never selected drop out of the model. As each base-learner is fitted separately, models with more covariates than observations can be fit and computational effort scales linearly in the number of covariate effects. By appropriately decomposing terms into e.g. a linear part and a nonlinear deviation, we cannot only select covariates, but also distinguish linear effects from smooth effects depending on the same covariate \citep[compare][]{KneibHothornTutz2009} and covariate interactions from additive marginal effects (See Online Supplement A.2).

	\section{Analysis of bacterial interaction in \textit{Escherichia coli} }
\label{chap_Application}

The coexistence of various bacterial species is a key factor in environmental systems. Equilibria in this biodiversity stand or fall with the species' interaction. Mathematically, bacterial growth can be considered as a stochastic process. A comprehensive analysis of this process involves investigating both mean and variability of growth curves. 

Certain bacteria strains produce toxins and use them to assert themselves in bacterial competition. \citet{vonBronk2016} establish an experimental setup with two cohabiting \textit{Escherichia coli} bacteria strains: a 'C-strain' producing the toxin ColicinE2 and a colicin sensitive 'S-strain' pipetted together on an agar surface. Single bacteria of the C-strain population sacrifice themselves in order to liberate colicin. The emitted colicin diffuses through the agar and kills numerous S-strain bacteria on contact. On the other hand, the S-strain might outgrow the C-strain and starts in a favored position of an initial ratio S:C of about 100:1. The arising population dynamics are influenced by external stress induced with the antibiotic agent Mitomycin C (MitC). MitC slightly damages the DNA of the bacteria. While it has little effect on the S-strain, it triggers colicin production in the C-strain as an SOS-response. A higher dose of MitC increases the fraction of colicin producing C-bacteria and, thus, colicin emission \citep{vonBronk2016}.
 
At a total of $N=334$ observation sites, bacteria under consideration are exposed to one of four different MitC concentrations. Bacterial growth curves $S_i(t)$ of the S-strain and $C_i(s)$ of the C-strain, $i=1,...,N$, are observed over 48 hours. Their values correspond to the propagation areas of the bacterial strains, which are obtained from the automated image segmentation procedure implemented by \citet{vonBronk2016}. S- and C-strain areas can be distinguished as the bacteria are marked with red and green fluorescence, respectively. The resulting area growth curves are measured on a fixed time grid with $G=105$ measurements per curve. The experiments are conducted in batches of about $40$ bacterial spots and with two batches for each MitC concentration. In order to keep track of the growing bacteria, the zoom level of the microscope was adjusted after $12\nicefrac{1}{4}h$, $18\nicefrac{1}{2}h$ and $33\nicefrac{1}{2}h$. As the performance of the automatic bacterial area segmentation may depend on the zoom level, it has to be incorporated into the analysis.

\subsection{Model for S-strain growth}
\label{section_applied_model}

In order to obtain insights into bacterial interaction dynamics, we model the $i$-th propagation area curve of the S-strain $S_i(t)$ in dependence on the C-strain growth and other covariates. While usually $S_i(t)>0$, it might equal zero, if the S-strain is completely extinct or masked by the fluorescence of the C-strain. Therefore, we assume a conditional zero-adapted gamma distribution for $S_i(t)$: the gamma distribution has a positive support and offers the flexibility to model both a location and a shape parameter conditional on the survival of the S-strain at time $t$. While there are different possible parameterizations, we model the mean $\mu_i(t)$ and the shape parameter $\sigma_i(t)/\mu_i(t)$ with $\sigma_i(t)$ the standard deviation, conditional on $S_i(t)>0$. In addition, we model the probability of extinction of the S-strain, $\extinctprop_i(t) = P\left(S_i(t) = 0\right)$ over time. Each component of the resulting parameter vector $\bvartheta_i(t) = \left( \vartheta_i^{(\mu)}(t), \vartheta_i^{(\nicefrac{\sigma}{\mu})}(t), \vartheta_i^{(\extinctprop)}(t) \right)^\T = \left( \mu_i(t), \frac{\sigma_i(t)}{\mu_i(t)}, \extinctprop_i(t) \right)^\T$ is modeled as
\begin{align*}
g^{(q)} \left(\vartheta^{(q)}_i(t)\right) &= \beta_0^{(q)}(t) + \beta_{MitC_i}^{(q)}(t) + \beta_{Batch_i}^{(q)}(t) + h_1^{(q)}(C_i\,,t) + h_2^{(q)}(C_i',t) \\ 
\intertext{for $q\in\{\mu, \nicefrac{\sigma}{\mu}, \extinctprop\}$, with link-functions $g^{(\mu)} = g^{(\nicefrac{\sigma}{\mu})} = \log$ and $g^{(\extinctprop)} = \text{logit}$, and with historical effects \citep{BrockhausGreven2016a}}
h_j^{(q)}(C_i,t) &= \int_{0}^{t} C_i(s) \beta_j^{(q)}(s,t) \, ds\, .
\end{align*} 

For each distribution parameter, the model includes a functional intercept $\beta_0^{(q)}(t)$. As there are only four MitC concentrations employed, they are considered as categorical grouping variable and represented by group specific intercepts $\beta_{MitC_i}^{(q)}(t)$ per MitC level centered around the functional intercept. As a functional random intercept, we include an additional group specific intercept $\beta_{Batch_i}^{(q)}(t)$ to compensate for batch effects, which are centered around $\beta_{MitC_i}^{(q)}(t)$ in order to preserve identifiability of the functional intercept. The impact of the C-strain (history) on S-strain growth is modeled using historical effects with coefficient functions $\beta_j^{(q)}(s,t)$. As for the current C-strain propagation $C_i(s)$, a historical effect for its derivative $C'_i(s)$ is included corresponding to the C-strain growth. C-strain bacteria sacrifice themselves in order to emit colicin and its production is costly by itself, so $C_i'(s)$ also reflects the colicin emission. The covariate curves are centered around their empirical point-wise mean curve, such that $\frac{1}{N} \sum_{i=1}^{N} C_i(s) = 0$ for each $s$, and scaled with the corresponding standard deviation, such that $\sd{C(s)} = 1$. For $C_i'(s)$ correspondingly. Doing so, the coefficient functions can be uniformly interpreted over the whole time span. By integrating, the historical effect includes information about the curves from time point $t=0$ to the current time point $t$. 
For mean and scale parameter, we employ a log-link, such that applying a second order difference penalty for the functional intercepts penalizes deviations from exponential growth.
All effect functions are modeled with cubic P-splines and 2nd order difference penalties. For the functional intercepts, we include an additional step-function base-learner to capture the different zoom levels applied during the experiment at fixed known time points.  
For the MitC and batch effects a ridge type penalty over factor levels is utilized to achieve the same number of effective degrees of freedom for all base-learners. A common step-length of $\steplen=0.1$ is used for $\mu$, $\sigma/\mu$ and $p$. We fit the model with both implemented GAMLSS boosting methods and decide for the 'non-cyclic' method, described in Section \ref{section_fit}, which performed better in 10-fold curve-wise bootstrapping and is computationally more efficient. 

\subsection{Results}
\label{section_results} 

\paragraph{MitC effect and effect of experimental batches}

\begin{figure}[H]
	\includegraphics[page = 1, width = .7\textwidth]{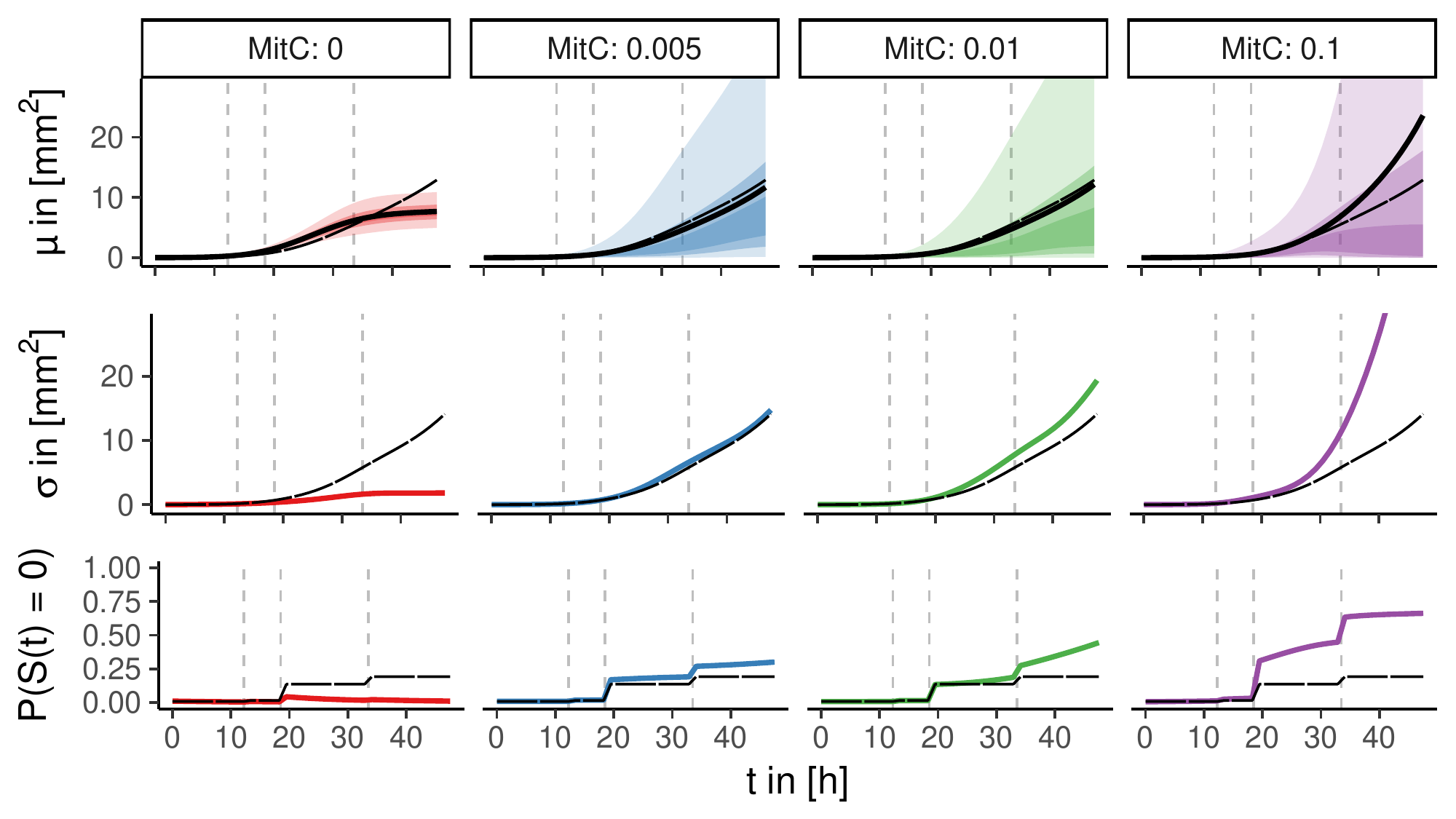}
	\centering
	\caption{ Estimated point-wise mean \textit{(top)} and standard deviation \textit{(center)} of the S-strain growth curves $S_i(t)$ conditional on $S_i(t)>0$ and the extinction probabilities \textit{(bottom)} of S-strain growth curves for each MitC concentration. Long-dashed black curves correspond to the functional intercept, dashed vertical lines to the zoom level change-points. Thick solid lines indicate the estimates, transparent ribbons reflect the point-wise inner 25\%, 50\% and 90\% probability mass intervals of the estimated gamma distributions conditional on $S_i(t)>0$ \textit{(top)} .} 
	\label{MitCeffects}
\end{figure}

An overview of the effects of the toxin MitC can be found in Figure \ref{MitCeffects}. We observe that mean S-strain growth is slightly increasing for low MitC levels compared to no MitC, but is particularly higher for $MitC_i = 0.1\, \nicefrac{\mu g}{ml}$. This indicates that, if $S_i(t) \geq 0$, the S-strain even grows better under this condition.
\newline
For the standard deviation, we observe a gradual but distinct rise with the MitC level. Due to the log-link we may not only interpret effects on the shape parameter $\nicefrac{\sigma}{\mu}$ but also on $\sigma$: effect functions $h_j^{(\sigma)}$ for $\sigma$ are obtained as $h_j^{(\sigma)} = h_j^{(\mu)} + h_j^{(\nicefrac{\sigma}{\mu})}$. In this plot, we choose to depict $\sigma$ instead of $\nicefrac{\sigma}{\mu}$, as it is more straightforward to interpret on the response level.\newline 
In order to visualize the combined effect on the response gamma distribution conditional on $S_i(t)>0$, the inner 25\%, 50\% and 90\% probability mass intervals of the estimated point-wise distribution are depicted as ribbons of different transparency. We observe that positive skewness increases with MitC concentration. \newline
It is important to note, that control experiments indicate no considerable effect of MitC on S-strain growth \citep{vonBronk2016}. Thus, present covariate effects of MitC might reflect effects of C-cells which cannot be explained by the observed C-strain growth curves. \newline
Showing distinct shifts at the zoom points, $\extinctprop_i(t) = P\left(S_i(t)=0\right)$ seems to depend highly on the zoom level of the microscope. This suggests, that besides full extinction of the S-strain, $S_i(t)=0$ is also linked to limitations in area recognition. 
Additionally, the probability for $S_i(t)=0$ is higher for positive MitC concentrations. Overall, the conditional mean for positive $S_i(t)$, but also the variability and probability for zero increase with the MitC concentration.
\newline
The smooth functional effects for each of the eight experimental batches are relatively small in size. For the conditional mean $\mu$ they cause an average deviation of about $3\%$ of the intercept growth curve (geometric mean over observed time points and batches); for the scale parameter $\nicefrac{\sigma}{\mu}$ the average deviation is about $9\%$; and for $p$ about $6\%$. 
While point-wise 95\% bootstrap confidence interval type uncertainty bounds (Online Supplement Section D.3) 
show less accuracy for the batch effects (in particular, those on $p(t)$), they indicate a high estimation precision for the MitC-effects and functional intercepts. This corresponds to our findings in the simulation study in Section \ref{section_simulation}. 

\paragraph{C-strain effect}

\begin{figure}
	\centering
	\includegraphics[width = 0.8\textwidth]{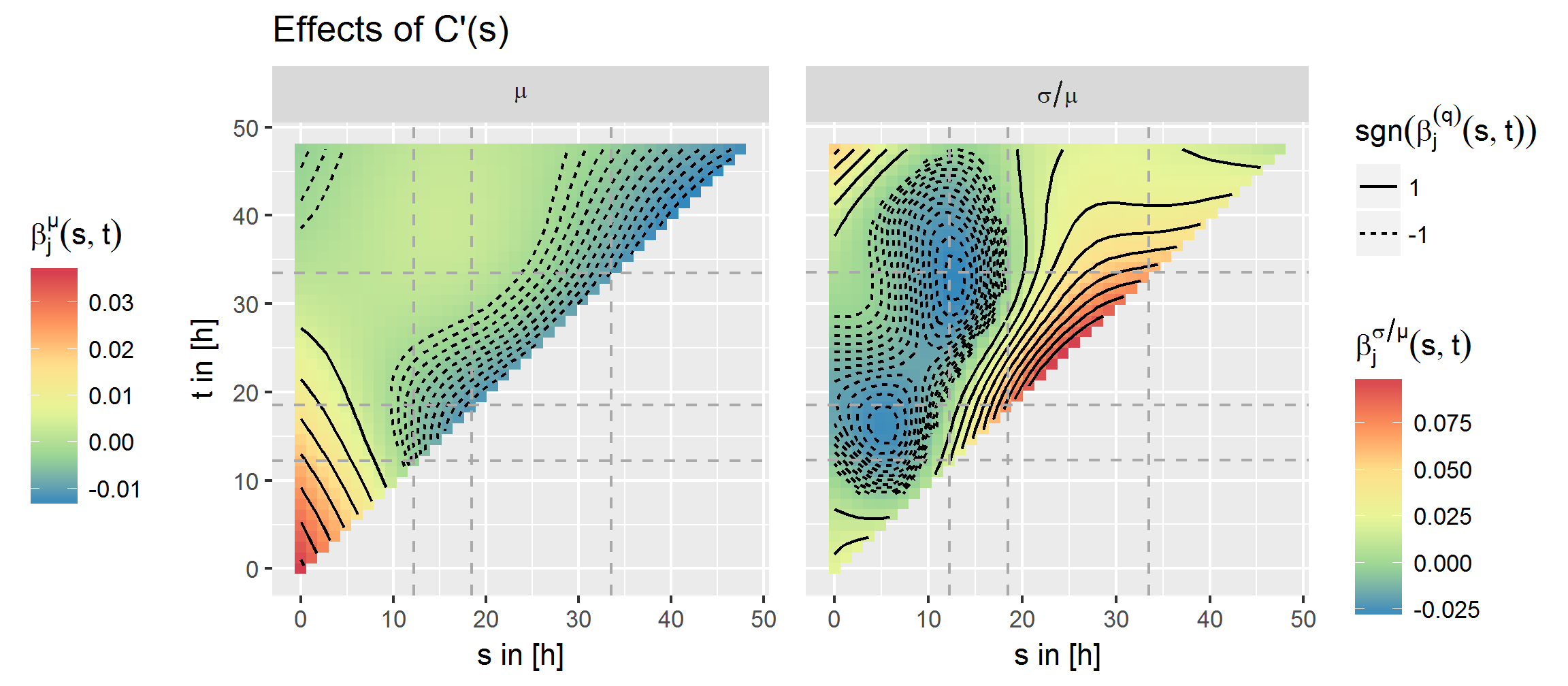}
	\caption{\textit{Left:} Coefficient function $\beta^{(\mu)}(s,t)$ for the historical effects of $C'_i(s)$ on the mean of S-strain growth curves. \textit{Right:} the corresponding plot for the effect of $C'_i(s)$ on the scale parameter $\nicefrac{\sigma_i(t)}{\mu_i(t)}$. The $y$-axis represents the time line for the response curve, the $x$-axis the one for the C-strain growth curve. The change-points in zoom level are marked with dashed gray lines. For a fixed $s=s_0$, $\beta^{(\mu)}(s_0,t)$ and $\beta^{(\sigma/\mu)}(s_0,t)$ describe the effect of the normalized covariate at time $s_0$ on the S-strain growth curve over the whole remaining time interval.}
	\label{dCdteffects}
\end{figure}

The base-learner for the C-strain area propagation $C_i(s)$ effect on $\mu_i(t)$ is never selected throughout the boosting procedure and the effect on $\nicefrac{\sigma_i(t)}{\mu_i(t)}$ is small (Online Supplement Figure 18). 
Thus, we only discuss the effect of the area increment $C'_i(s)$ here (Figure \ref{dCdteffects}). 
Looking at the $C'$-$\mu$-effect (effect of $C'(s)$ on mean S area), we can distinguish two main impact phases. 

In the earlier growth phase with $s \leq 10\, h$, we observe a positive $C'$-$\mu$-effect concerning almost the whole time curve of the S-strain. That means that C-strain growth above [below] the average indicates increased [decreased] S-strain propagation. Both colicin production and colicin secretion are costly to the population and slow down C-propagation. A low value of $C_i'(s)$ indicates early colicin secretion. We conclude that this first phase delineates a time window, where colicin emission is able to severely harm the S-strain population.
\newline
In the second phase for $s > 10\, h$, we observe a negative $C'$-$\mu$-effect, which is maximal at short time lags and slowly fading. This likely reflects spatial competition of the S- and the C-strain (compare Online Supplement Figure 11).
At this time, bacteria have grown together to coherent formations and strains obstruct expansion of each other. 
\newline
Even though the $C'$-$\mu$-effect offers this clear interpretation, it is rather small compared to, e.g., the MitC effect on $\mu(t)$. Moreover, while in simulation studies we observe a rather high estimation precision for most of the historical effects (Figure \ref{appSim_rel_RMSE}), 95\% bootstrap confidence interval type uncertainty bounds indicate distinctly less estimation precision than for the MitC-effects (Online Supplement Figures 17,18). 
However, the historical $C'$-$\nicefrac{\sigma}{\mu}$-effect also corroborates the distinction into two phases of interaction: while in the first phase there is a negative effect of C-strain growth, the effect turns positive in the second phase. 
Thus, relative variability is increased for slow C-strain growth early in the experiment (colicin production) and for fast C-strain growth later in the experiment (areal competition).

For the probability $\extinctprop_i(t) = P(S_i(t)=0)$ both the effects of $C_i(s)$ and $C'_i(s)$ were selected. Corresponding plots can be found in the Online Supplement Figure 15).
However, as already indicated by the marked cuts between the different zoom levels (Figure \ref{MitCeffects}), vanishing of the S-strain is particularly sensitive to the precision of the area recognition. Hence, we are careful with interpreting the effects. Moreover, as for $t \leq 10\, h$ the probability $P(S_i(t)=0)\approx0$ and there is almost no vanishing of $S$ in the data, the estimated effects in this period are questionable and, thus, we refrain from interpreting them. For later growth periods we observe a mainly positive effect of $C_i(s)$ and $C'_i(s)$ on $p(t)$ corresponding to an increased probability for $S_i(t)=0$ for an increased C strain growth (areal competition), consistent with results for $\mu(t)$. Only the $C'$-$\mu$-effect for $10\, h < s < 20\, h$ is estimated negative for later S-strain growth.
Summarizing our results in this section, we observe effects of the C-strain on the S-strain growth curves on all three target distributional parameters which are accompanied by effects of the MitC concentration. The inherent model selection yields that -- given a positive area of the S-strain -- not the C-strain propagation $C_i(s)$, but rather the growing behavior, given by the derivative $C'_i(s)$, presents the relevant factor for the response mean curves; and considering the shape of the historical functional effect of $C'_i(s)$, we can distinguish similar time phases of bacterial interaction as \cite{vonBronk2016}. We note that $\extinctprop(t)$ highly depends on questions of area recognition and, thus, focus our substantive interpretation on the effects on $\mu(t)$ and $\nicefrac{\sigma}{\mu}(t)$. This is also motivated by the fact that killed S-strain bacteria do not immediately vanish but rather their area stops growing. While we, thus, concentrate on the model part conditioning on $S_i(t)>0$,  we observed qualitatively similar effects of $C'(s)$ on the mean, when replacing zero response values by the smallest observed positive values in preliminary analyses. This indicates that $C'(s)$ shows a similar effect on the unconditional mean $\E[S_i(t)] = \mu_i(t)(1-\extinctprop_i(t))$.\\ 
Based on the estimates for $\extinctprop(t)$ and for $\mu(t)$ and $\nicefrac{\sigma}{\mu}(t)$ conditioning on $S(t)>0$, we may also compute the unconditional mean and standard deviation of $S(t)$, e.g., for the different MitC concentrations (Online Supplement Figure 14).
Doing so, we observe that the MitC concentration only has a minor effect on the unconditional mean, though still a large effect on the unconditional variance, and, hence, its influence on the S-strain growth could not be captured by non-GAMLSS regression models for the mean only. 
	\section{Simulation studies}
\label{chap_Simulation}

Although related boosting approaches to non-functional or one-parameter special cases of the present model are well tested (e.g. see \citet{BrockhausGreven2016b}, \citet{BrockhausGreven2015}, \citet{Thomas2016}, \citet{MayrSchmid2012}), estimating GAMLSS for functional response naturally poses additional challenges. In several simulation studies, we confirm that gradient boosting presents a suitable estimation approach. In particular, we observe that cross-validation on the curve-level is desirably sensitive to in-curve dependency of response measurements and, therefore, prevents over-fitting. As this is crucial for this novel approach, we concentrate on the capability of the presented method to yield good estimates despite the model complexity and in-curve dependency and refer to \citet{Thomas2016} for a discussion of the variable selection quality of gradient boosting for GAMLSS.

We perform two different simulation studies: the first is a large simulation study for the case of Gaussian functional response curves, where we model both mean and standard deviation in dependence of scalar and categorical covariates. The second simulation study is motivated by the bacterial interaction model applied in Section \ref{chap_Application}. It includes functional covariates and response measurements following a zero-adapted-Gamma distribution. In both studies, we explore the fitting behavior with respect to different tuning parameters, sample sizes and in-curve auto-correlation structures. The overall goodness-of-fit is evaluated in terms of the Kullback-Leibler-Divergence (KLD); the estimation quality is assessed with the root mean square error (RMSE) (for further details see Online Supplement C.1).


\subsection{Gaussian response model simulation}
\label{section_genericSim}

As in scalar regression, Gaussian functional response models play a prominent role in functional response regression. However, in contrast to previous regression frameworks, we consider the case where both mean and standard deviation depend on covariates. In this large scale simulation study, we consider two different models: one with scalar continuous covariates $z_1, z_2 \in [0,1]$ and one with categorical covariates $g_1, g_2 \in \{1,...,4\}$. The covariates $z_1$ and $g_1$ influence both the mean $\mu(t)$ and the standard deviation $\sigma(t)$. $z_2$ and $g_2$ influence only the mean. The continuous covariate model includes a covariate interaction for $\mu(t)$. Precise model formulations are presented in Table \ref{sim_models}.

\begin{table}
	\caption{Simulated models with scalar continuous and categorical covariates \label{sim_models}}
	\small
	\centering
	\begin{tabular}{l l}
		\hline
		\textbf{Models:} & \hfill Distribution: $ Y(t)\,|\,\bx \sim \mathcal{N}\left( \mu\left(t\right), \sigma^2\left(t\right) \right) $ \\ \hline
		\textbf{1. Continuous} & $ \mu(t) = f_0^{\mu}(t) + f_1^{\mu}(z_1, t) + f_2^{\mu}(z_2, t) + f_3^{\mu}(z_1, z_2, t)$ \\
		& $ \log{\sigma(t)} = f_0^{\sigma}(t) + f_1^{\sigma}(z_1, t) $ \\
		\textbf{2. Categorical} & $ \mu(t) =\beta_0^{\mu}(t) + \beta_{g_1}^{\mu}(t) + \beta_{g_2}^{\mu}(t)$ \\
		& $ \log{\sigma(t)} = \beta_0^{\sigma}(t) + \beta_{g_1}^{\sigma}(t) $ \\
		\hline
	\end{tabular}
\end{table}

We simulate smooth covariate effect functions $f\jq, \beta\jq$ generating them as cubic B-splines with random coefficients. Note that the smooth interaction effect $f_3^{\mu}(z_1, z_2, t)$ is fairly complex as it involves a double tensor product spline basis, and we distinguish this interaction effect from the marginal effects $f_1^{\mu}(z_1, t)$ and $f_2^{\mu}(z_2, t)$ using basis-orthogonalization. 

\begin{figure}
	\centering
	\foreach \x in {1,2,3}
	{
		\includegraphics[page = \x, width = .24\linewidth]{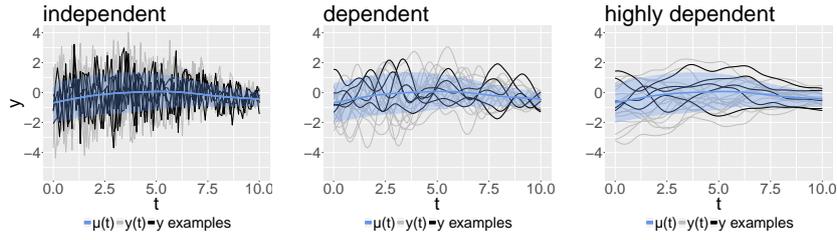}
	}
	
	\caption{Sampled response curves for different in-curve dependency levels.
		A sample of N = 20 response curves (four highlighted in black) sampled around the mean curve \textit{(blue)} over time for fixed covariate values. The blue area plotted around the mean corresponds to $\pm$ the standard deviation. Three levels of in-curve dependency are presented in increasing order. Compare Online Supplement B.}
	\label{dependency_examples} 
\end{figure}

The auto-correlation of the response curves is induced by function specific random Gaussian error processes $\varepsilon(t)$, such that $Y(t) = \mu(t) + \sigma(t)\varepsilon(t)$ with $\varepsilon(t) \sim \mathcal{N}(0, 1)$. We distinguish three different levels of auto-correlation: an independent case, where $\varepsilon(t)$ is independent identically distributed white noise, and a dependent and a highly dependent case. For the latter two, we generate $\varepsilon(t) = \mathbf{b}^\top(t)\boldsymbol{\zeta} / \sqrt{\mathbf{b}^\top(t)\mathbf{b}(t)}$ by standardizing splines with basis vector $\mathbf{b}(t)$ and standard normal random coefficients $\boldsymbol{\zeta}$. Doing so, we obtain smooth response curves, as shown in Figure \ref{dependency_examples}. For further details see Online Supplement B.

\begin{figure}
	\centering
	\includegraphics[page = 4, width = \textwidth]{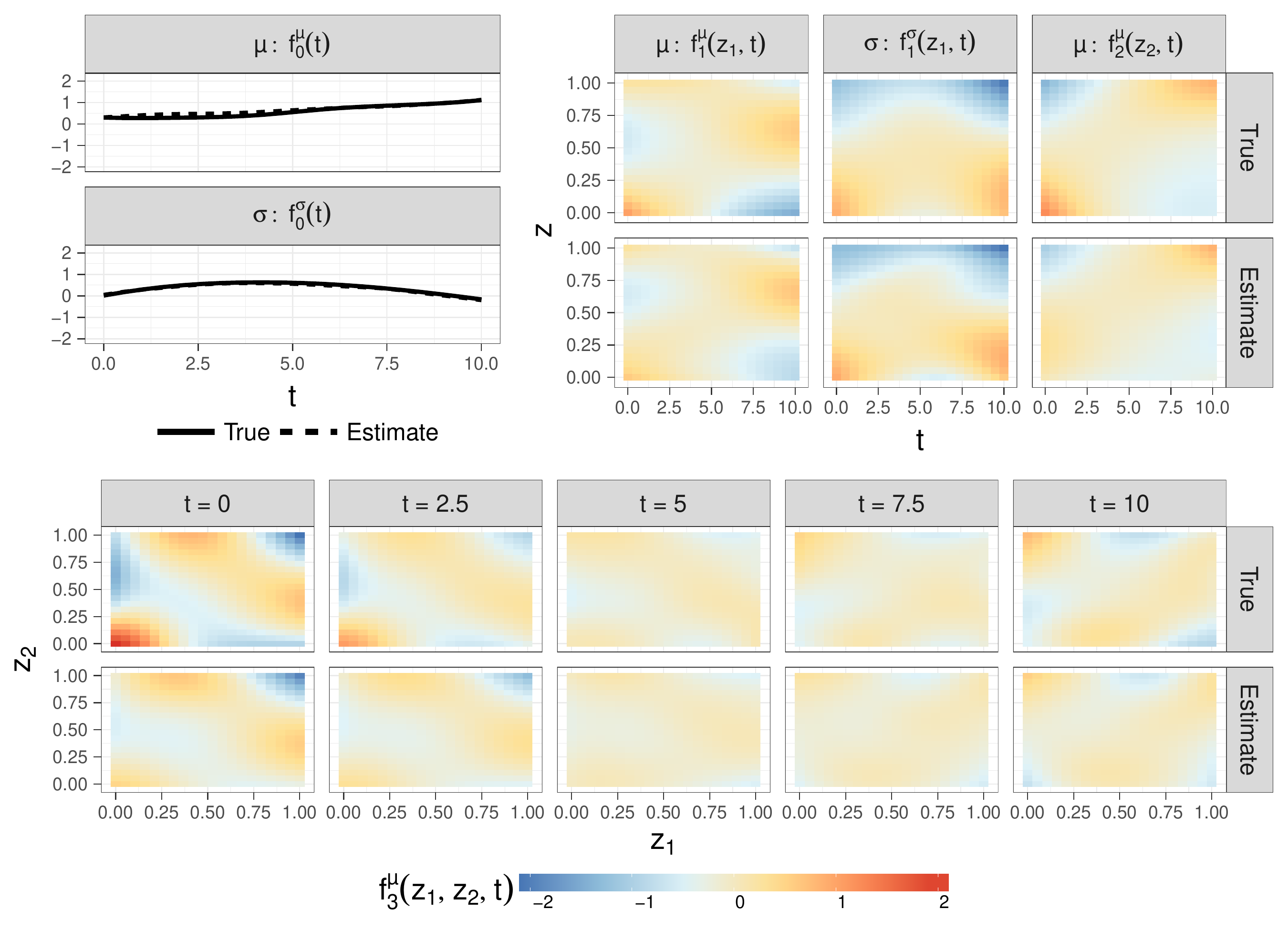}
	
	\caption{Example fit for high response in-curve dependency.
		The figure shows an example of original and estimated effect functions for the simulated continuous covariate model scenario. The fit is based on a moderate number of $N = 334$ response-curves with $G = 100$ highly dependent measurements per curve and respective covariate samples. It is representative, in the sence that it corresponds to the median $\overline{\KL}$ from the true simulated model. As the covariate interaction depends on $z_1,\ z_2$ and $t$, its effect functions are plotted for five fixed time points \textit{(bottom)}.}
	\label{fit_example}
\end{figure}

In order to achieve comprehensive results, we consider a series of different simulation scenarios, where we vary the following parameters: the model scenario, i.e. continuous or categorical covariates; the sample size $N$ and the number of measurements per curve $G$; the amount of in-curve dependency in the three categories described above; the boosting and re-sampling method; the effective degrees of freedom of the base-learners and the step length $\nu$; and the ratio of the variance of the randomly generated true predictors for the mean and for the standard deviation. In addition, we compare the performance with the penalized likelihood approach of \cite{GrevenScheipl2017}. An overview can be found in Online Supplement C.4.
In the following, the model fitting behavior will be summarized from different perspectives.

\begin{figure}
	\includegraphics[width = \textwidth]{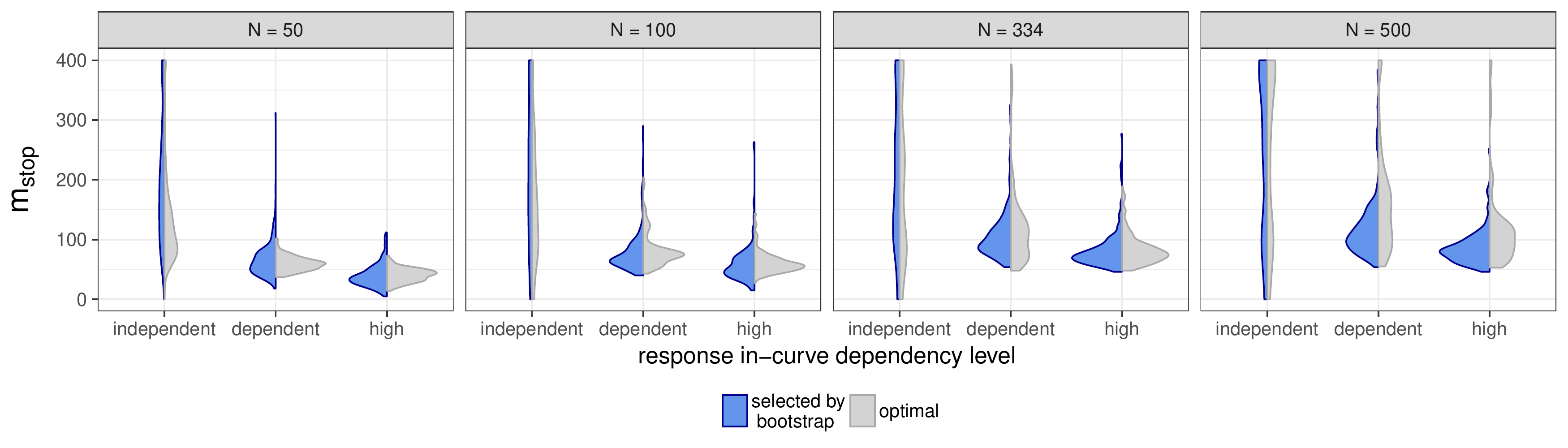}
	\includegraphics[width = \textwidth]{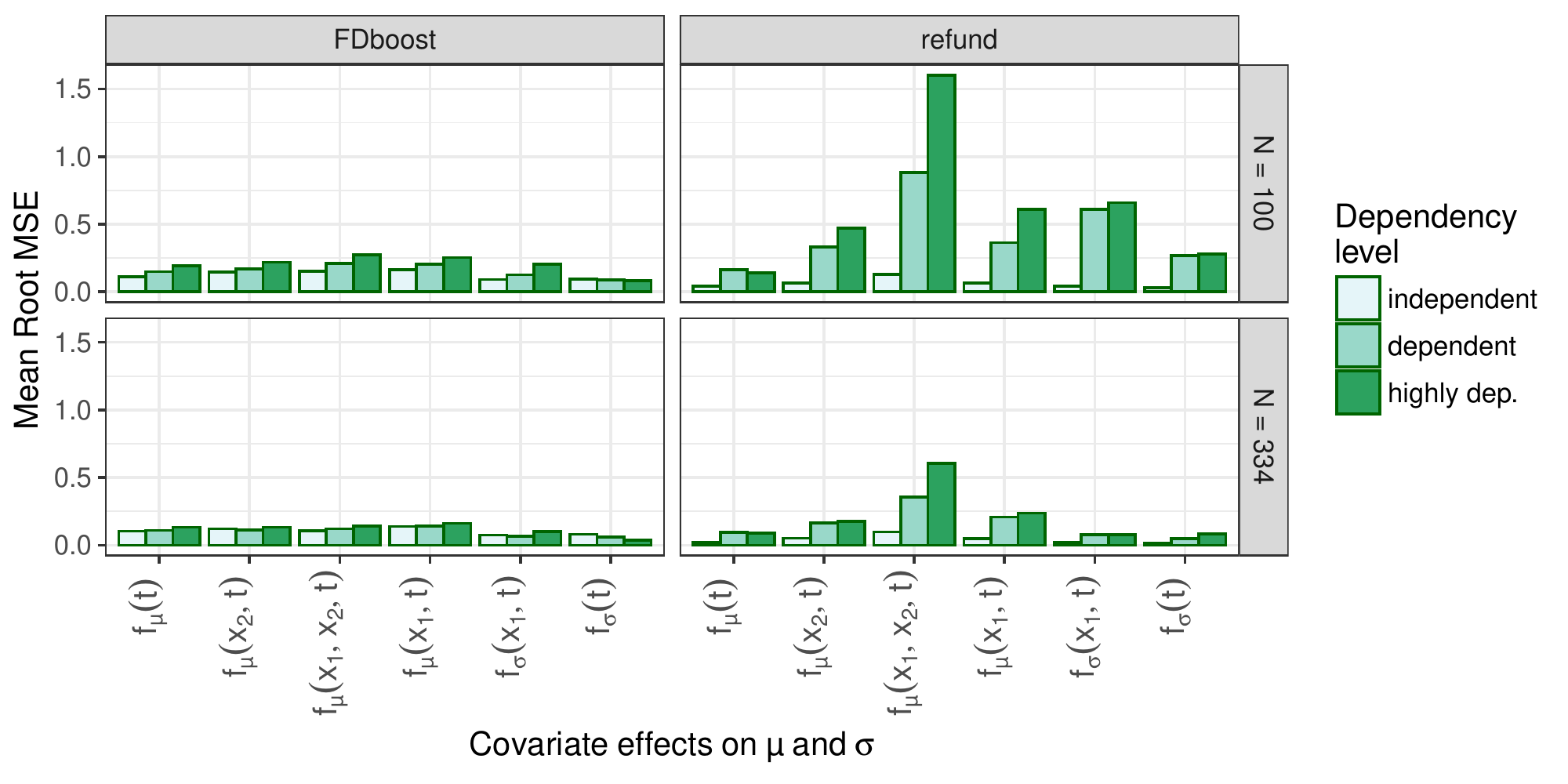}
	
	\centering
	
	\footnotesize
	\caption{
		Plots referring to the more complex model scenario with continuous covariates. For each combination of sample size $N$ and in-curve dependency level, 10 true predictors and 20 sets of error processes $\epsilon(t)$ were sampled, yielding 200 data sets for each combination. 
		\textit{Top:} 
		Violin-plots reflecting the empirical distribution of the stopping iterations $m_{stop}$ selected via 10-fold bootstrap (blue; left) and for the $\overline{\KL}$-optimal $m_{stop}$ (gray; right). Note that, as gradients decrease in absolute size with the number of boosting iterations, earlier iterations have a greater effect on the actual model fit than later iterations.
		\textit{Bottom:}
		Bar-plots indicating the mean RMSE of the different effects for data scenarios with $N=100$ and $N=334$ functional observations for our approach based on gradient boosting (\textit{left}) and the approach based on penalized likelihood (\textit{right}).
	}
	\label{mstop}
	
\end{figure}

\paragraph{Preventing over-fitting with early stopping} Especially for a small number $N$ of sampled curves, high in-curve dependency promotes over-fitting. Flexible time-varying effect functions, as entailed in the present models, might be fitted to random patterns occurring in single curves. Early stopping in gradient boosting solves this problem. For dependent response measurements the optimal stopping iteration $m_{stop}$, with respect to the mean Kullback-Leibler Divergence ($\overline{\KL}$) over the domain, is much lower than for the independent case. At the same time, a larger sample size allows for more fitting iterations. As shown in Figure \ref{mstop}, this is captured well by estimating the $\overline{\KL}$-optimal $m_{stop}$ with 10-fold bootstrap on the curve-level. 
For high in-curve dependency bootstrapping or sub-sampling turn out to perform better than cross-validation, which tends to over-estimate the $\overline{\KL}$-optimal $m_{stop}$ (Online Supplement Figure 6).\\ 
The simulation study shows that early stopping plays a key role in gradient boosting for functional GAMLSS. As the information gain per measurement of a response curve is diminished by in-curve dependency, it is crucial that the available information is not over-estimated. In the case of in-curve dependency early stopping greatly improves the model fit, as otherwise over-fitting might easily occur (see also Online Supplement Figure 8).\\ 
This becomes even clearer when comparing the fitting performance with the GAMLSS-type regression for Gaussian response presented by \cite{GrevenScheipl2017}, which is fit with the \texttt{R} package \texttt{refund}. In this alternative approach, the model is fit via penalized maximum likelihood based on a working independence assumption for the response curve measurements. Usually, independence is assumed conditionally on a latent Gaussian random error process, which is included into the model to account for in-curve dependency. However, this would prohibit modeling the total variance of the response curves in a separate predictor, and, thus, no process is included here. In contrast to the early stopping in our \texttt{FDboost} boosting approach, the current version of \texttt{refund} has no mechanism to account for in-curve dependency in the case of GAMLSS. The simulation results in Figure \ref{mstop} show that this may lead to severe over-fitting. While for independent response measurements the effect estimates are even slightly more accurate with \texttt{refund}, the fit is far better with \texttt{FDboost} for response curves with realistically dependent measurements. This is particularly visible for the smooth covariate interaction effect, the most complex effect in the model. The over-fitting in \texttt{refund} in this case is clear when comparing single example model fits (Online Supplement Figure 3). 


\paragraph{Sample size} We compare sample sizes $N = 50, 100, 334, 500$ for a constant grid of $G = 100$ measurements per response curve. In the independent case, we obtain good estimations from $N = 50$ on (Online Supplement Figures 2 and 4). 
For dependent and highly dependent response measurements estimation is naturally harder due to the decrease in independent information provided. However, despite the high complexity of the model, we observe that the structure of all effect functions is re-covered fairly well also in the highly dependent case (Figure \ref{fit_example}).
Here, the performance considerably improves with increasing sample size, such that we obtain good results for $N=100$.
In addition, the model fit gets more stable with higher sample sizes showing less variation of estimation quality across samples. \\
When comparing grid sizes $G = 20, 50, 100, 150$ for $N = 100$ response curves we observe no remarkable differences in estimation quality except for the independent case (Online Supplement Figure 5).
The in-curve dependency structure reduces the `effective number of measurements' provided by a fine grid. \\
Corresponding in some sense to the signal-to-noise ratio of simpler simulation scenarios, we control the scales of predictors for the mean and for the standard deviation by specifying the variances of the randomly generated true predictors. In most of the simulation scenarios both variances are chosen to be one. If the scale of the mean predictor is doubled, we observe that overall fitting gets worse with respect to the KLD. In this case, we observe an increased estimation quality for effects on the mean, but much worse estimation of the effects on the variance (Online Supplement Figure 2 and 4). 

\paragraph{Estimation of effects} For the estimation quality of the individual effect functions, we obtain similar results as discussed above for the global estimation quality. As expected, the fitting error tends to increase with the effect complexity. In addition, we observe that the covariate effects of variables which affect both mean and variance show an increased fitting error. However, neither mean nor variance effects show a clear general fitting advantage (see Figure \ref{mstop} and for more detail Online Supplement Figure 4).


\subsection{Application-motivated simulation}
\label{section_simulation}

The second simulation study is motivated by the analysis of bacterial growth in Section \ref{chap_Application}. The model and covariates perfectly correspond to those of Section \ref{section_applied_model}, which involves a zero-adapted gamma distribution for the response and, i.a., functional historical covariate effects on three distributional parameters $\bvartheta(t) = \left( \mu(t), \frac{\sigma(t)}{\mu(t)}, \extinctprop(t) \right)^\T$. Consequently, the model estimated from the data in Section \ref{section_applied_model} is taken as the true underlying model in the simulations.\\
The functional response $Y(t)$ is generated for three levels of in-curve dependency: 'independent', 'dependent' and 'highly dependent', which are induced by the same smooth Gaussian processes $\varepsilon(t)\sim\mathcal{N}(0,1)$ as in Section \ref{section_genericSim}. However, now we use inverse transform sampling by setting $Y(t) = F_{\bvartheta(t)}^{-1}(\Phi(\varepsilon(t)))$, where $F_{\bvartheta(t)}^{-1}$ denotes the quantile function of the zero-adapted gamma distribution with the parameters $\bvartheta(t)$ and $\Phi$ is the standard Gaussian CDF. Doing so, we obtain continuous functional observations in the dependent and highly dependent case following the desired distribution (Online Supplement B.2.2).\\ 
For each dependency level 120 simulation runs are performed. 
Comparing the bacteria growth curves from Section \ref{chap_Application} with simulated response curves, we observe that their in-curve dependency might be roughly comparable to the 'highly dependent' simulation scenario (Online Supplement C.3).

Concerning the comparison of different dependency levels and the estimation of the optimal stopping iteration, we obtain similar results to those of Section \ref{section_genericSim}: accuracy and stability decrease with increasing dependency and selection of the best stopping iteration $m_{stop}$ with cross-validation is sensitive to in-curve dependency. In this scenario, we observe a tendency of over-shooting the $\overline{\KL}$-optimal $m_{stop}$ (Online Supplement Figure 10).
Still, for all levels of in-curve dependency, the $\overline{\KL}$ for the optimal $m_{stop}$ and the $\overline{\KL}$ achieved with the bootstrap-selected $m_{stop}$ are quite similar, showing that the discrepancy in the stopping iteration has little impact on the fitting quality.\\

\begin{figure}
	\centering
	\includegraphics[page = 2, width = \textwidth]{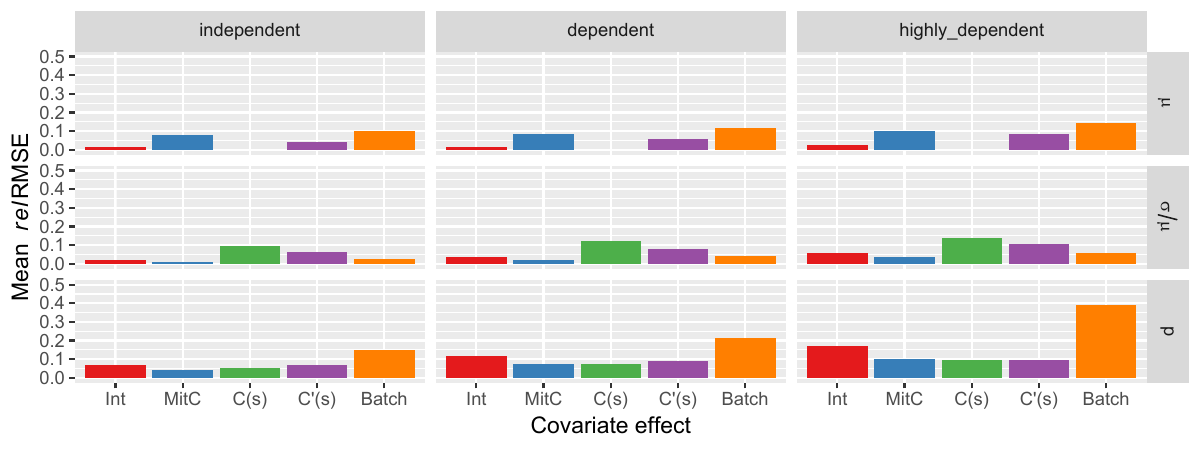}
	
	\caption{\textit{rel}RMSE per covariate effect: 
	We compare the mean \textit{rel}RMSE over 120 simulation runs per covariate effect in the model, and per in-curve dependency level. Effect functions are depicted for the mean $\mu$ and for the relative standard deviation $\nicefrac{\sigma}{\nu}$, as well as for the zero-probability $p$. The bars for the $C$-$\mu$-effects are missing, as -- after not being selected by the boosting algorithm in the original model -- their relative error is not defined. For a similar plot of the plain RMSE see Online Supplement Figure 10.}
	\label{appSim_rel_RMSE}
\end{figure}

Figure \ref{appSim_rel_RMSE} shows a comparison of the mean RMSE relative to the range of the particular effect (\textit{rel}RMSE) for the estimated effects in the simulation. 
We observe that most of the RMSEs are lower than 10\% of the effect range even in the highly dependent setting. However, the functional intercept in the predictor for $p(t)$ appears to be harder to estimate, being composed of a smooth functional intercept and a step function; the $\nicefrac{\sigma}{\mu}$-effect of $C(t)$ has a relatively large \textit{rel}RMSE, due to its small effect size, while having a quite small absolute RMSE (Online Supplement Figure 10); 
the group-effects for the eight experimental batches show the largest \textit{rel}RMSEs. They also have comparably small effect sizes and the estimated effect for each batch may rely on the respective data only. Still, for the batch effects the interpretation of their exact shape is not of primary interest. \\ 
Overall, we observe the effects to be estimated quite well despite in-curve dependency and the high complexity of the model.

\section{Discussion and outlook}
\label{chap_Discussion}

The functional GAMLSS regression framework we present in this paper allows for very flexible modeling of functional responses. We may simultaneously model multiple parameters of functional response distributions in dependence of time and covariates, specifying a separate additive predictor for each parameter function. In addition, point-wise distributions for the response curves beyond exponential family distributions can be specified. Doing so a vast variety of new data scenarios can be modeled and
 these new possibilities have shown to be crucial, when applying the framework to analyze growth curves in the present bacterial interaction scenario. To account for the complexity of the matter, we have to assume a three-parameter zero-adapted gamma distribution for the measurements of the S-strain growth curves, when modeling them in dependence of the C-strain and covariates.
 
The proposed regression framework is a generalization of earlier scalar-on-function GAMLSS and functional additive models for function-on-function or function-on-scalar regression as summarized by \cite{GrevenScheipl2017}, which in turn generalize scalar GAMLSS, GAM, generalized linear or linear regression models.
While the new GAMLSS for functional responses is the first to achieve all requirements of the presented application, extending the existing model framework and software allows us to specify 
the smooth functional group effects, step function effects, and historical functional effects 
necessary to grasp the nature of the data scenario.


The results we obtain confirm and extend previous work: 
Focusing on the outcome after 48h and on the number of C-clusters at the edge of the S-colony after 12h, \citet{vonBronk2016} already distinguish two phases of bacterial interaction similar to those we find in the historical functional effects of the C-strain growth. The functional regression model not only provides new evidence for this distinction from a completely new perspective, but now also allows to quantitatively discuss the effect over the whole time range. Moreover, now we observe C-growth to have a positive effect on S-growth in the early phase and a negative effect in the later phase. The separation of these two phases appears even more distinct in the effect on the relative standard deviation, which we would not be able to recognize without GAMLSS.

Regarding the fraction of S- and C-strain area after 48h \citet{vonBronk2016} categorized three different states of the bacterial interaction: for no MitC, there is either dominance of the S-strain or coexistence; for a moderate MitC concentration, there occurs a splitting into two extremes -- either dominance of the S-strain or extinction; and for the highest MitC concentration, the toxin strategy of the C-strain fails and the S-strain either dominates or both strains go extinct. Now, referring to the complete growth curves, our results also reflect this categorization: if MitC is added, and conditional on a positive area, the mean S-strain growth increases, whereas also the probability for zero area and the variance increase. However, these differences would not be captured by non-GAMLSS regression models for the mean only, as the unconditional mean growth curves for no and moderate MitC concentration are very similar (see Online Supplement Figure 14). 
Apart from that, the framework provides the flexibility to account for special challenges in the experimental setup, such as dependencies between observations in the same experimental batch and differences between zoom levels of the microscope.\\

In simulation studies we confirm that by fitting our models via component-wise gradient boosting we are capable of estimating even complex covariate effects on multiple distributional parameter functions. As it prevents over-fitting, early stopping of the boosting algorithm based on curve-wise re-sampling plays a key role: it enables us to face settings with highly auto-correlated response curves.\\ 
 
Still, it is a dedicated aim of future research to explicitly incorporate the auto-correlation of the response curves into the framework. A popular approach is to do this with curve-specific functional random intercepts \citep{Morris2015}. 
While our framework includes (curve-specific) functional random intercepts, there is, yet, no sensible way to incorporate them into the mean structure without interference with the estimation of other distributional parameters in the GAMLSS. We thus did not include a curve-specific functional random intercept in our models, to let the GAMLSS model capture all the variation in the data conditional on covariate effects.\\
Further, we want to extend the presented ideas to multivariate functional responses. This is also relevant for an extension of the present discussion of bacterial interaction, when we consider experimental setups with more than two bacterial strains involved.
	\section*{Acknowledgements}
\vspace{-.2cm}
Financial support from the Deutsche Forschungsgemeinschaft (DFG) through Emmy
Noether grant GR 3793/1-1 (AS, SB, SG) and through grant OP252/4-2 part of the DFG Priority Program SPP1617 is gratefully acknowledged. B.v.B was supported by a DFG Fellowship through the Graduate School of Quantitative Biosciences Munich (QBM). Additional financial support by the Center for Nanoscience (CeNS) and the Nano Systems Initiative - Munich (NIM) is gratefully acknowledged. 
	
	  \bibliographystyle{chicago}
	  \bibliography{literatur}

 \newpage
 \appendix
 %
%
%
%
%
%

\renewcommand\thefigure{\thesection.\arabic{figure}}    
\setcounter{figure}{0}   

\textbf{\large Web-based supporting materials for Boosting Functional Response Models for Location, Scale and Shape with an Application to Bacterial Competition by St\"ocker, Brockhaus, Schaffer, Bronk, Opitz and Greven}
%

This Online Supplement contains supplementary material in four different respects: Appendix \ref{app_basis} provides more details concerning the construction of partial effect functions from tensor product bases, as discussed in Section 2 of the main manuscript; Appendix 
\ref{app_sim_details_1} introduces the technical background necessary to draw appropriate smooth random response curves and random effect functions in the GAMLSS-scenario for both simulation studies presented in Section 4; Appendix \ref{app_sim_details_2} gives more detailed insights into the setup of the simulation studies, the used measures of estimation quality, and the obtained results; and Appendix \ref{app_app} contains additional information concerning the analysis of bacterial interaction in Section 3, i.e. data details, a comparison to other growth models in the literature and further model results.

\section{Basis representations and orthogonalization}
\label{app_basis}

\subsection{Tensor product bases}
\label{app_tensor}

For both fitting and prediction, it is necessary to evaluate effect functions $h_j$, as defined in Section 2.1, at all data points simultaneously. Following \citet{Scheipl2015}, this can be written in a simple closed form expression using row tensor products.

\paragraph{Definiton} Consider an $n\times m$ matrix $\bA$ and an $r\times s$ matrix $\bB$. The \emph{Kronecker product} '$\otimes$' is defined as
\begin{equation*}
		\bA \otimes \bB = 
		\begin{bmatrix}
		a_{1,1}B	& \dots	 & a_{1,m}B      \\
		\vdots 	& \ddots & \vdots 			\\
		a_{n,1}B 	& \dots & a_{n,m}B  	 \\
		\end{bmatrix}
\end{equation*}

	If $\bA$ and $\bB$ have the same number of rows $n = r$, the \emph{row tensor-product} '$\odot$' is defined by
\begin{equation*}
		\bA \odot \bB = 
		\begin{bmatrix}
		\ba_1 \otimes \bb_1	\\
		\vdots 	\\
		\ba_n \otimes \bb_n 	\\
		\end{bmatrix}	\ ,	
\end{equation*}

where $\ba_i$ and $\bb_i$ are the $i$-th rows of the matrices $\bA$ and $\bB$.

Consider two variables $t\in T$ and $\bx\in \mathcal{X}$. Let $ h_j(\bx, t) = ( \bb_{Xj}(\bx,t) \otimes \bb_{Yj}(t))^\T \btheta_j \hspace{1cm} $ with parameter vector $\btheta_j$ and function basis vectors $\bb_{Xj}(\bx, t)$ and $\bb_{Yj}(t)$ as defined in Section 2.1. For a data set with $N$ observations and $G$ measurements per response curve, let covariates $\bX = \vec{\bx_{1,1}, \dots, \bx_{1,G}, \bx_{2,1}, \dots, \bx_{N,G}}$ and respective time points $\bt = \vec{t_{1,1}, \dots, t_{1,G}, t_{2,1}, \dots, t_{N,G}}$. Define 
\begin{equation*}
	\bB_{Xj}(\bX, \bt) = 
	\begin{bmatrix}
	\bb^\T_{Xj}(\bx_{1,1},t_{1,1})	\\ 
	\vdots 	\\
	\bb^\T_{Xj}(\bx_{N,G},t_{N,G})	\\
	\end{bmatrix}		 
\end{equation*}

the $NG\times K_j$ design matrix. Analogously, define the vector $\bH_j(\bX,\bt)$ of length $NG$ to entail the evaluations of $h_j$ and the $NG \times K_Y$ design matrix $\bB_{Yj}(\bt)$ with the evaluations of $\bb_{Yj}$. Then, we can write the joint evaluation as equation
\begin{equation}
\label{rowTensor}
	\bH_j(\bX,\bt) = \big( \bB_{Xj}(\bX, \bt) \odot \bB_{Yj}(\bt) \big)^\T \btheta_j
\end{equation}
in terms of the row tensor product. In the same way, covariate effect bases of two different covariates can be combined to an interaction effect. Note that a common number of measurements per curve $G$ is 
only assumed for ease of notation. 

If all response curves are observed on a common grid $\bt_0 = \vec{t_1,...,t_G}$ and covariates and their effect bases are not time dependent, equation \eqref{rowTensor} can be re-written in terms of the Kronecker product. In this case, we can denote covariates as $\bX_0 = \vec{\bx_{1}, \dots, \bx_{N}}$. Then the covariate part simplifies to an $N\times K_j$ matrix $\bB_{Xj}(\bX_0) = \bB_{Xj}(\bX_0, \bt_0)$ that is independent of $\bt_0$ and also the $\bB_{Yj}(\bt_0)$ can be arranged as a $G\times K_Y$ matrix. For these matrices, we obtain 
\begin{equation}
\label{Kronecker}
\bH_j(\bX,\bt) = \big( \bB_{Xj}(\bX_0) \otimes \bB_{Yj}(\bt_0) \big)^\T \btheta_j \, .
\end{equation}
The Kronecker product has desirable mathematical properties, which simplify the implementation of linear constraints in Section \ref{app_constraints}. In particular, it enables us to consider the model as Functional Linear Array Model \citep{BrockhausGreven2015}, which increases computational efficiency. In terms of Generalized Linear Array Models \citep{Currie2006}, the model can be fitted without actually computing the 
Kronecker product. 

\subsection{Othogonalization of effect functions}
\label{app_constraints}
Consider two effect functions $h_1(\bx, t) = \bb^\T_1(\bx, t) \btheta_1$ and  $h_2(\bx, t) = \bb^\T_2(\bx, t) \btheta_2$ with function bases $\bb_{1}$ and $\bb_{2}$ of dimension $K_1$ and $K_2$. Assume constant functions are entailed in $span(\bb_2)$, i.e. $\exists \btheta_c:\bb_2(\bx, t) \btheta_c = 1\ \forall\bx,t$. Assume the same for $\bb_1$. This property typically holds for, e.g., spline bases. We want to construct an interaction effect in terms of the Kronecker product basis $\bb_1(\bx, t)\otimes \bb_2(\bx, t)$. Applying $\btheta_c$ from above we get
\begin{equation*}
h_1(\bx, t) = \bb_1^\T(\bx, t)\btheta_1 = \bb_1^\T(\bx, t)\btheta_1 \otimes \bb_2^\T(\bx, t) \btheta_c = \left(\bb_1(\bx, t)\otimes \bb_2(\bx, t)\right)^\T \left(\btheta_1 \otimes \btheta_c \right) \, .
\end{equation*}
This is analogously obtained for $h_2$. Hence, $h_1, h_2 \in span(\bb_1\otimes\bb_2)$. However, for separate model fitting, interpretation and automatic model selection, we want to include marginal effects $h_1, h_2$ and their interaction $h_{12}$ as distinct effects into the additive predictor. To do so, we have to construct linearly independent design matrices:

For observed covariates $\bX$ and $NG$ time points $\bt$, let $\bB_1 = \bB_1(\bX, \bt)$ and $\bB_2 = \bB_2(\bX, \bt)$ be the $NG\times K_j$ design matrices as defined in Section \ref{app_tensor}. The $k$-th column of $\bB_1$ corresponds to the evaluations of the $k$-th basis function $b_{1,k}(\bx, t)$. The same for $\bB_2$. Let $\widetilde{\bB} = \bB_1 \odot \bB_2$ denote the complete tensor product design matrix. Then, we obtain the design matrix $\bB$ for the interaction effect via a linear transform $\bB = \widetilde{\bB}\bZ$. The transformation matrix $\bZ$ is specified as a matrix with $NG$ columns and with a maximum number of orthogonal rows such that $\bB^\T\bB_1 = \bB^\T\bB_2 = \boldsymbol{0}$. This means, we construct $\bB$ such that it is orthogonal to the design matrices of the marginal effects. Applying QR-decomposition to the matrix $\bC = \widetilde{\bB}^\T \big[\bB_1 : \bB_2\big] $ a suitable matrix $\bZ$ is determined by
\begin{equation*}
	\bC = \begin{bmatrix}
	\bQ : \bZ \\
	\end{bmatrix}
	\begin{bmatrix}
	\, \bR \, \\
	\, \boldsymbol{0} \,\\
	\end{bmatrix} = \bQ\bR \, ,
\end{equation*}

since $\bB^\T \big[\bB_1 : \bB_2\big] = \bZ^\T\widetilde{\bB}^\T \big[\bB_1 : \bB_2\big] = \bZ^\T \bQ \bR = \boldsymbol{0}$, i.e. $\bB$ is orthogonal to $\bB_1$ and $\bB_2$ \citep{Wood2006, BrockhausGreven2015}. According to \citet{BrockhausGreven2015}, we proceed just the same way in order to distinguish effect functions $ h_j(\bx, t) = (\bb_j(\bx,t) \otimes \bb_Y(t))^\T \btheta_j$ from the functional linear intercept $h_0(\bx, t) = \bb^\T_Y(t) \btheta_0$, which is a special case of the above. Computation can be further simplified in the linear array case.

\section{Simulation details I:\\ generation of random smooth errors and effect functions}
\label{app_sim_details_1}

In order to sample appropriate data for a simulation study concerning functional GAMLSS, we have to draw appropriate smooth random response curves and, at the same time, we do not only have to flexibly control the mean, but also the variance and other distributional parameters over time. In the Gaussian response simulation study, we additionally sample the true underlying effect functions to increase representativeness. For both purposes, we have to control the smoothness of the randomly drawn curves in order to achieve realistic samples. This section introduces the technical framework used to implement these points.

\subsection{Random spline generation}
\label{app_rbs}
The simulation relies crucially on random spline generation. Given a closed interval $T \subset \R$, a random spline $r:T\rightarrow\R,\ t \mapsto \bb^\T(t)\btheta$ is understood as the product of a fixed spline basis vector $\bb(t) = \vec{b_1(t),\dots,b_K(t)}$ and a random vector $\btheta$ of $K$ coefficients. In the following, $\btheta$ is always $\mathcal{N}(\zero, \bI_K)$ distributed, with any necessary transformation subsumed into the basis $\bb(t)$, i.e. $\bb(t)$ is obtained by transforming an underlying prototype basis $\tilde{\bb}(t) = \vec{\tilde{b}_1(t), \dots, \tilde{b}_K(t)}$. For $\tilde{\bb}(t)$ we take a B-spline basis of degree $l$ with $K - l + 1$ equally spaced knots. The random splines are scaled employing a scale parameter $\bar{\sigma}^2\geq 0$. In correspondence to P-splines, a $d$-th order difference penalty analogue can be specified, which is controlled with a smoothing parameter $\lambdaslash \in [0,1]$. This is accomplished by specifying the random spline as $r(t) = \bb^\T(t)\btheta = \tilde{\bb}^\T(t)\bOmega\bW \btheta$ with a suitable $K\times K$ orthogonal matrix $\bOmega$ and a diagonal weight matrix $\bW$ depending on two parameters $\bar{\sigma}^2$ and $\lambdaslash$, which will be further clarified below. 

In order to adjust the degree of smoothness of randomly drawn splines, we follow the mixed model representation of P-splines \citep{Fahrmeir2004}. Let $\bP = \bD^\T\bD$ be a $d$-th order difference penalty matrix inducing a quadratic penalty of the form $\tilde{\btheta}^\T\bP \tilde{\btheta}$, with $d\leq l$. Then there is a quadratic orthogonal matrix $\bOmega$, such that for $\tilde{\btheta} = \bOmega\btheta$ we obtain $\tilde{\btheta}^\T\bP \tilde{\btheta} = ( \bOmega \btheta)^\T\bP \bOmega \btheta = \btheta^\T \bI_K^{(d)} \btheta$, where $\bI_K^{(d)}$ is the diagonal matrix with the first $d$ diagonal entries zero and the rest one. Thus, by multiplying with $\bOmega^\T$ the basis $\tilde{\bb}(t)$ is transformed such that $b_{1}(t), \dots, b_{d}(t)$ represent the unpenalized part of the spline and $b_{d+1}(t), \dots, b_{K}(t)$ are subject to a ridge penalty. A suitable transformation matrix is given by $\bOmega = \big[ \bL : \bD^\T (\bD \bD^\T)^{-1} \big]$. The $m$-th column of the $K\times d$ matrix $\bL$ is given by $\bL_m = \vec{p_{m-1}(1),\dots, p_{m-1}(K)}$ with orthogonal polynomials $p_0,...,p_{d-1}$ of order $0,...,d-1$, such that $\bL^\T\bL = \bI_d$. In the mixed model estimation, the penalized coefficients $\theta_{d+1},...,\theta_K$ are considered random, whereas the unpenalized coefficients $\theta_d,...,\theta_K$ are considered fixed. However, for drawing the whole vector $\btheta$, this would lead to a improper distribution. So, we also add some variance to the unpenalized coefficients and control the trade-off between the variance for the penalized and unpenalized parts with the smoothing parameter $\lambdaslash$, which corresponds to the part of the variance of the randomly generated spline curve, which is explained by the unpenalized smooth part. Moreover, we want to be able to control the scale of the randomly generated spline curves and, therefore, use the parameter $\bar{\sigma}^2$, which is given by the total variance of the random curves. These two parameters determine $\bW$. More precisely, they are defined as follows:

Evaluating the prototype B-spline basis $\tilde{\bb}(t)$ on a grid $\bt\in T^G$ we obtain a 'design' matrix $\tilde{\bB}$ with $\tilde{\bb}^\T(t_1), \dots, \tilde{\bb}^\T(t_G)$ as its rows.
For a given $\bW$, $\bB = \tilde{\bB}\bOmega\bW$ presents the corresponding matrix for the desired spline basis $\bb(t)$ of the random spline. Drawing $\btheta\sim \mathcal{N}(\zero, \bI_K)$, the covariance matrix for the random spline evaluations $\br = \bB\btheta$ on $\bt$ is given by $\Cov(\br) = \bB\bB^\T$. For the scale parameter we employ the mean variance over time $\bar{\sigma}^2 = \overline{\Var}(\br) = \frac{1}{G} \tr(\bB\bB^\T) = \frac{1}{G} \tr(\bB^\T\bB)$. To control the smoothness of the resulting random spline curves we use $\lambdaslash = 1 - \frac{\tr(\bB^\T\bB \bI_K^{(d)})}{\tr(\bB^\T\bB)}$, which corresponds to the percentage of $\overline{\Var}(\br)$ explained by the unpenalized part of the spline. It presents a natural parameter of smoothness: Similar to the smoothing parameter $\lambda$ in penalized regression, a high value of $\lambdaslash$ corresponds to a high degree of smoothness, because, in this case, most of the variance in the randomly drawn spline curves stems from the smooth unpenalized part. However, $\lambdaslash$ and $\lambda$ are not directly mathematically related.\\ 
Technically, the above definitions of $\lambdaslash$ and $\bar{\sigma}^2$ are implemented via specifying the weight matrix $\bW^2 = \bar{\sigma}^2 \left( \frac{\lambdaslash}{d\bar{\sigma}^2_{un}} (\bI_K - \bI_K^{(d)}) + \frac{(1-\lambdaslash )}{(K-d)\bar{\sigma}^2_{pe}} (\bI_K^{(d)}) \right) $ with $\bar{\sigma}^2_{un} = \frac{1}{G} \tr\left((\tilde{\bB}\bOmega)^\T\tilde{\bB}\bOmega (\bI_K - \bI_K^{(d)})\right)$ the mean variance of the unpenalized part and $\bar{\sigma}^2_{pe} = \frac{1}{G} \tr\left((\tilde{\bB}\bOmega)^\T\tilde{\bB}\bOmega \bI_K^{(d)}\right)$ the mean variance of the penalized part.

Tensor product random splines depending on multiple variables are generated and orthogonalized as described in Sections \ref{app_tensor} and \ref{app_constraints}. For tensor product random splines the total mean variance is adjusted in a further step.
	
\subsection{Sampling smooth response curves}
\label{app_incurvedependency}
	
	In order to sample response curves with the desired smoothness, i.e. the desired in-curve dependency structure, we sample splines with random coefficients, and then transform them, such that they follow the desired point-wise distribution of the response. 
	\subsubsection{The Gaussian case} 
	We start with the Gaussian case, where we may re-formulate the functional GAMLSS models as
	\begin{equation}
	\label{model1}
	y_i(t) = \mu(\bx_i, t) + \gamma_i(t) + \varepsilon_{i,t}
	\end{equation} 
	with $t\in T$, covariates $\bx_i$, a mean structure $\mu(\bx_i, t)$, a smooth random error curve $\gamma_i(t)$ and independent errors $\varepsilon_{i,t}\sim\mathcal{N}(0, \sigma_\varepsilon^2(\bx_i, t))$, with the joint error $\gamma_i(t) + \varepsilon_{i,t}\sim \mathcal{N}(0, \sigma^2(\bx_i, t))$. In order to sample smooth error curves with the desired properties, we represent them as $\gamma_i(t) = w(\bx_i,t) r_i(t)$ with $w(\bx_i,t)$ a suitable weight function and 
	$ r_i(t) = \bb^\T(t) \bzeta_i $ a random spline
	with a vector of $K_Y$ basis functions $\bb(t)$ and $\bzeta_i \sim \mathcal{N}(\mathbf{0},\mathbf{I}_{K_Y})$ constructed as described above in Section \ref{app_rbs}.\\ 
	Hence, for obtaining the desired joint error variance it must hold that
	$$ \sigma^2(\bx_i, t) = (w(\bx_i, t))^2\cdot \bb^\T(t)\bb(t) + \sigma_\varepsilon^2(\bx_i, t)\, , $$
	as the variance of the random spline is given by $\Var (r_i(t)) = \bb^\T(t)\Cov(\bzeta_i) \bb(t) = \bb^\T(t) \mathbf{I_{K_Y}} \bb(t)$.
	This is achieved by setting $w(\bx_i, t) = \sqrt{\frac{\sigma^2(\bx, t)-\sigma_\varepsilon^2(\bx, t)}{\bb^\T(t)\bb(t)}}$.\\
	In our simulating studies, we either restrict to the smooth error $\gamma_i(t)$ for response curves with in-curve dependency or to $\varepsilon_{i,t}$ for the independent case.
	
%
%

	\subsubsection{The general case}
	\label{app_generate_curves_general}
	Let the random spline $r_i(t) = \bb^\T(t) \bzeta_i$ of degree $l$ be defined as above. Consider a probability distribution with a cumulative distribution function (CDF) $F$, such that its quantile function $F^{-1}$ is $l-1$ times continuously differentiable (of class $\mathcal{C}^{l-1}$). Let $F$ depend on $Q$ parameters in a parameter space $\Theta\subseteq\R^Q$ and assume $F^{-1}$ is also $\mathcal{C}^{l-1}$-differentiable with respect to these parameters. Let $\bvartheta_i: T \mapsto \Theta$ be an $\mathcal{C}^{l-1}$-differentiable parameter function and let $\Psi$ denote the CDF of the standard normal distribution. Then, 
	\begin{equation*}
	y_i(t) = F^{-1} \left(\; \Psi \left(\, \frac{r_i(t)}{\sqrt{\bb^\T(t)\bb(t)}}\, \right)\; \big| \bvartheta_i(t)\; \right)
	\end{equation*} 
	is a $\mathcal{C}^{l-1}$-differentiable curve and for every $t\in T$ the $y_i(t)$ is marginally distributed according to $F$ with the current parameter setting $\bvartheta_i(t)$. This is due to chain rule and inversion method (see, e.g., \citet{Devroye1986}).

\section{Simulation details II: simulation studies}
\label{app_sim_details_2}

\subsection{Measures for evaluation}
\label{app_measures}
In accordance with the fitting aim formulated in Section 2.2, the general goodness-of-fit of a model is measured with respect to the Kullback-Leibler divergence to the true probability distributions: we use $ \overline{\KL}(\boldsymbol{\hat{\bh}}) = \frac{1}{NG} \sum_{i=1}^{N} \sum_{t\in T_0} \KL\left[\mathcal{F}_{Y(t)|\bX} : \mathcal{\widehat{F}}_{Y(t)|\bX} \right] $, the mean Kullback-Leibler divergence over all point-wise evaluations for an estimated predictor $\boldsymbol{\hat{\bh}}$, where $\mathcal{F}_{Y(t)|\bX}$ is the true distribution with the true parameter functions $\btheta_i(t)$ and $\mathcal{\widehat{F}}_{Y(t)|\bX}$ is the one with the estimated $\hat{\btheta}_i(t)$ for the $i$-th curve.

In order to evaluate and compare the goodness-of-fit of the individual effects or coefficient functions, we rely on the root mean square error (RMSE). In the application motivated simulation study, where the scale of the true covariate effects are not controlled, the RMSE is normalized by the range of the particular true effect to achieve comparability. In this case, the applied individual goodness-of-fit measure is given by
\begin{align*}
	rel\text{RMSE}\left(\hat{f}\jq\right) &= \text{RMSE}\left(\hat{f}\jq\right) \big/ \left(\underset{t}{\max}\, f\jq(t) - \underset{t}{\min}\, f\jq(t)\right)  \hspace*{.5cm} \text{or}\\ 
	rel\text{RMSE}\left(\hat{\beta}\jq\right) &= \text{RMSE}\left(\hat{\beta}\jq\right) \big/ \left(\underset{s,t}{\max}\,\beta\jq(s,t) - \underset{s,t}{\min}\,\beta\jq(s,t) \right)\ ,  
\end{align*}
for effect or coefficient functions, respectively.

\subsection{Gaussian response model simulation}
\label{app_simulation}

This section provides details concerning the simulations discussed in Section 4. We simulate data for a functional response model with point-wise normally distributed measurements. Mean and standard deviation are modeled in dependence of continuous or categorical covariates, respectively. 

\subsubsection{Sampling}
\label{app_sampling}

The simulations follow a two-stage sampling approach: in each run, $n_{eff}$ true effect function sets and $n_{cov}$ covariate and random error sets are sampled. Evaluating the true effect functions on the covariates yields a total of $n_{eff}n_{cov}$ simulated data sets.

For $N$ observations, continuous or categorical covariates are drawn independently from a uniform distribution on the unit interval or on $\{1,2,3,4\}$, respectively. 

All effect functions are drawn as random splines (Section \ref{app_rbs}). The scale of mean and variance effects is specified via scale parameters $\bar{\sigma}_\mu^2,\; \bar{\sigma}_\sigma^2 > 0$. $\bar{\sigma}_\mu^2$ corresponds to the overall mean variance over time of the effect functions, i.e. a weighted sum of the individual random spline scale parameters. All effect functions except for the interaction effect are weighted equally. The interaction effect is weighted with $1/8$ as for this type of effect more extreme values occur. 
For $\bar{\sigma}_\sigma^2$, the overall mean variance of the effect functions is transformed for comparability, as a log-link is applied for the standard deviation: 
The overall mean variance $\tau^2 = \tau^2\left( \bar{\sigma}_\sigma^2 \right)$ is specified such that for a random variable $Z\sim \mathcal{N}(0,\tau^2)$ we obtain $\Var(\exp(Z))=\bar{\sigma}_\sigma^2$. 
Via the properties of the log-normal distribution, $\tau^2$ is determined as $\tau\left( \bar{\sigma}_\sigma^2 \right) = -\log(2) + \log( \sqrt{ 4 \bar{\sigma}_\sigma^2 + 1} + 1)$.  
If not otherwise specified, we set  $\bar{\sigma}_\mu^2 = \bar{\sigma}_\sigma^2 = 1$. 
All smooth effect function bases are chosen as cubic B-splines with 2nd order difference penalty using the smoothing parameter $\lambdaslash = 0.8$, which results in sensibly smooth true effect functions. For functional intercepts we use B-spline bases with $k = 8$ basis functions. The tensor product bases have a total of $k=3 \cdot 8= 24$ basis functions for categorical effect functions (4 categories minus reference times the size of the intercept basis), $k = 6\cdot 5 = 30$ for smooth marginal continuous effect functions (6 for time and covariate, respectively, minus 6 for orthogonalization constraint) and $k = 6\cdot 6\cdot 8 - 2\cdot30 -8 = 220$ for the smooth interaction of continuous covariates (6 for both covariates, 8 for the intercept, minus sizes of the other bases). All effects are centered with respect to the particular functional intercept and the continuous interaction effect is orthogonalized with respect to its marginal effect functions.

Given the covariate values and the true underlying effect functions, smooth response curves are drawn randomly with the respective mean and standard deviation given by the model. Each curve is evaluated on $G$ grid points of the time interval $T=[0,10]$. Point-wise, the response curves are normally distributed. As described in Section \ref{app_incurvedependency} smoothness is induced with random error splines. For obtaining different levels of in-curve dependency, different random splines are employed. For level \emph{independent}, we do not use any random splines, but sample directly from the marginal distribution of $y_i(t)$. For level \emph{dependent}, we use a cubic random spline with no smoothing penalty and $k=20$ basis functions. We use a first order difference penalty with $\lambdaslash=0.5$ for level \emph{high\_dependency}. 

\subsubsection{Model specification}
\label{app_simmodel}

Corresponding to sampled true effect functions, all scalar function bases used to construct the effect functions occurring in the fitted model are specified as cubic P-splines with second order difference penalty. All base learners are set up such that they have the same total degrees of freedom $\df$. The default is $\df=13$. To obtain this, an additional Ridge penalty is applied to the categorical effects. The P-spline bases for the functional intercepts contain $k=20$ basis splines, which is more than for the true intercepts such that the knowledge of the true knots is not used in the model fit. For other effects, the same basis dimension $k$ is used as for the corresponding true effect. Again, all effects are centered around the functional intercept and marginal scalar effects are distinguished from the interaction.

For the hyper-parameters a default of step-size $\nu = 0.2$ is employed and by default the optimal stopping iteration $m_{stop}$ is estimated by 10-fold curve-wise bootstrap. Furthermore, unless otherwise stated, the model is fitted with the GAMLSS boosting method described in Section 2.2, which corresponds to the 'noncyclic' boosting method in the R package \texttt{gamboostLSS}.

\subsection{Application-motivated simulation study}
\label{app_appsim}

In this simulation study, we adopt model and covariates entirely from the analysis of bacterial interaction in Section 3. Estimated effect functions from the original model fitted to the data present the true model structure in the simulation. Response curves are generated on the basis of random splines as described in Section \ref{app_incurvedependency} and with the same specifications as discussed in \ref{app_sampling} for the generic simulation study. Thus, in contrast to the above simulation study, we have one set of true effect functions and one set of covariates, only. In each simulation run new response curves are sampled. 
We evaluate the model on 120 data sets per dependency level and compare \emph{independent}, \emph{dependent} and \emph{high\_dependency} response curves. 

\begin{figure}[H]
	\centering
	\includegraphics[width=0.6\linewidth]{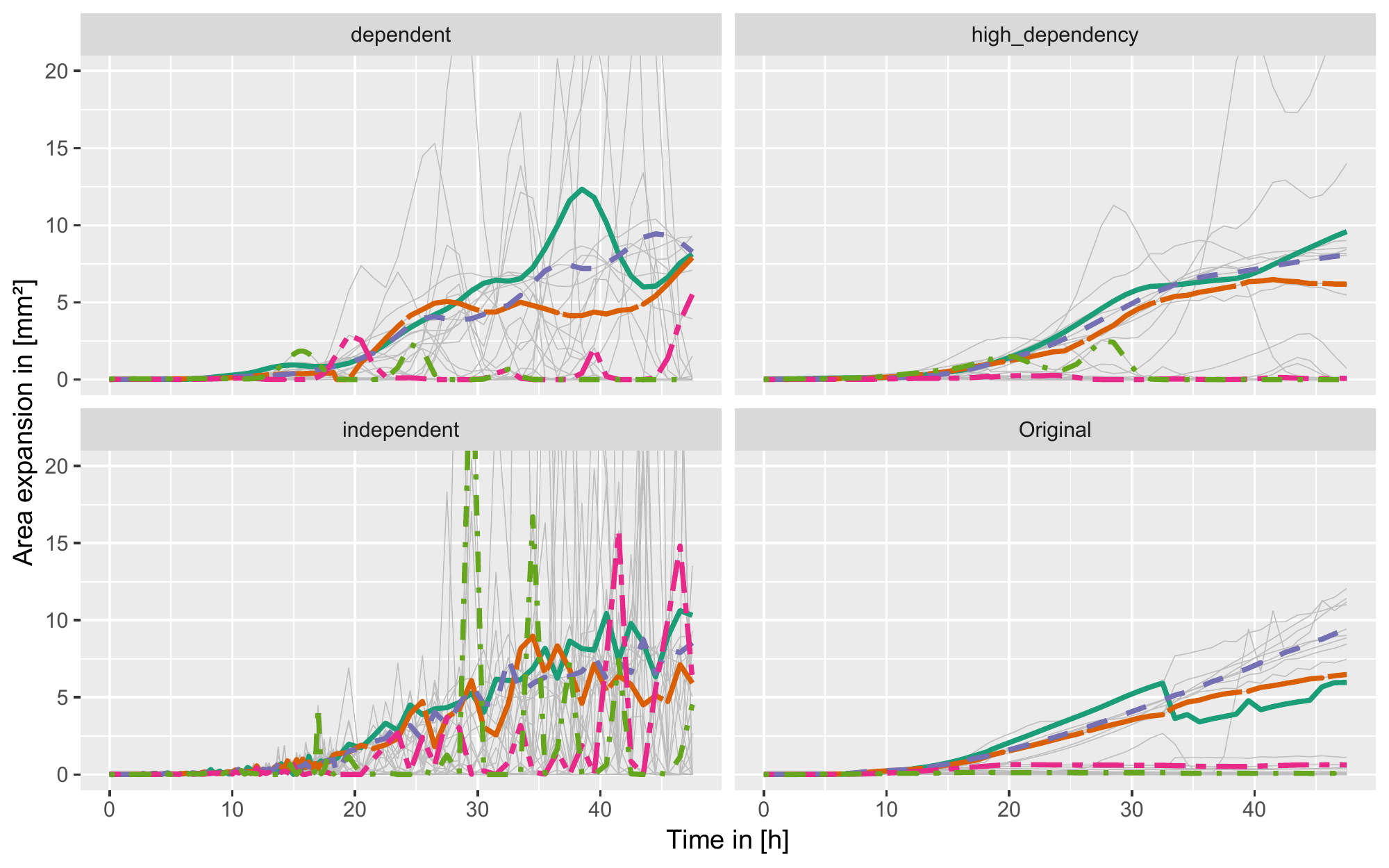}
	\caption{24 simulated example curves for the applied dependency levels \emph{independent}, \emph{dependent} and \emph{high\_dependency} and original growth curves for comparison. Five randomly selected curves are highlighted in colors. Corresponding curves have the same covariate setting.}
	\label{app_data_comparison}
\end{figure}

Figure \ref{app_data_comparison} shows simulated response curves for each level of in-curve dependency, as well as corresponding original growth curves for comparison.

Visual comparison of the simulated curves and the original growth curves suggests that \emph{high\_dependency} might be roughly comparable to the in-curve dependency in the data. However, as we do not model the dependency structure in the presented model, the correspondence is limited.

\subsection{Simulation overview}
\label{app_simOverview}

In order to provide both a comprehensive picture, comparing several simulation parameters, and a sound sample size per setting, multiple simulation runs are performed. Each focuses on different aspects of interest. Table \ref{table_simOverview} gives an overview on simulation runs and supplementary figures.

\begin{figure}[H]
	\centering
	\includegraphics[width = .76\textwidth]{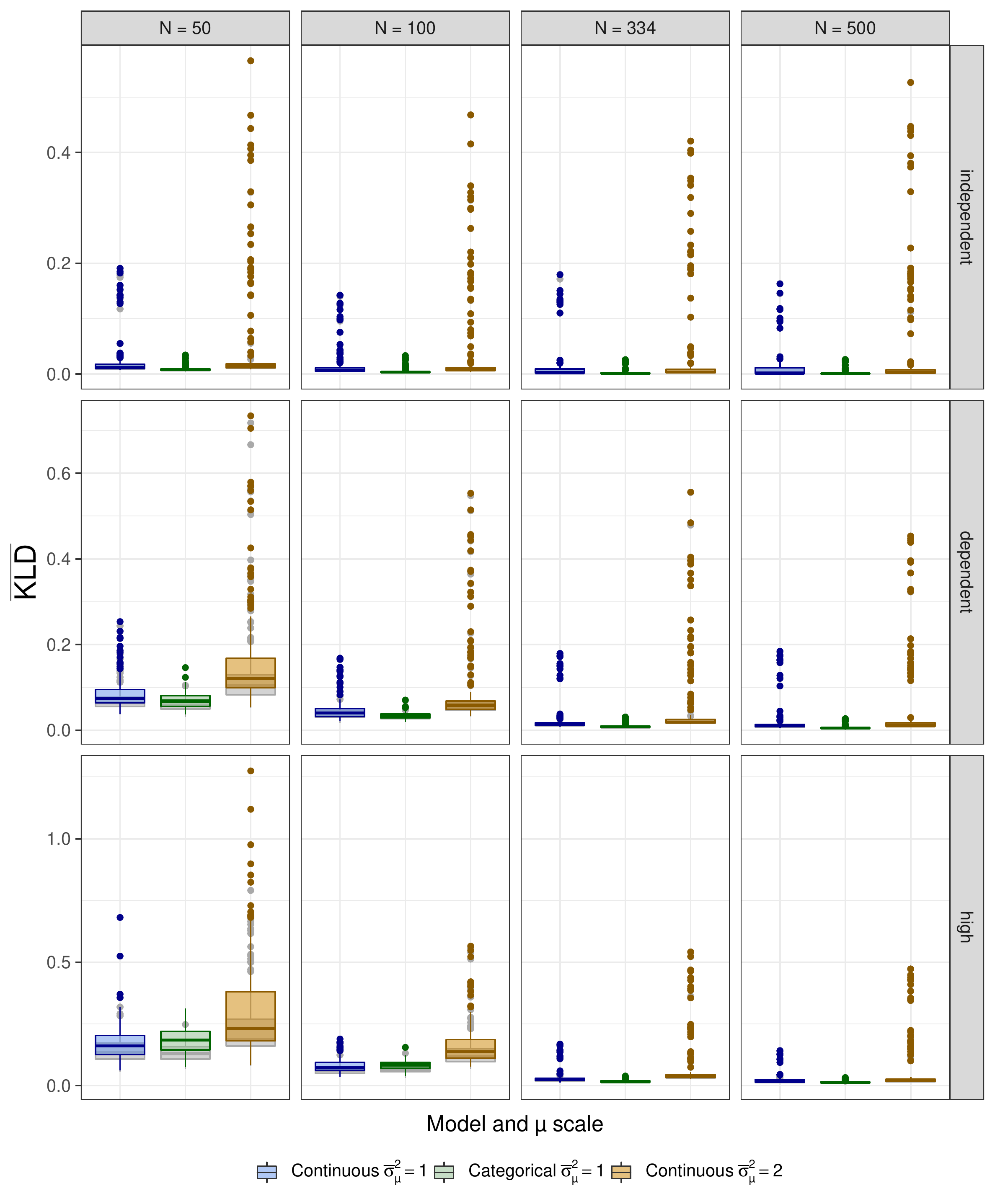}
	\caption{$\overline{\KL}$ for continuous and categorical covariate model:
	Comparison of $\overline{\KL}$ in the continuous models with $\bar{\sigma}_\mu^2 = 1, 2$ and for the categorical model with $\bar{\sigma}_\mu^2 = 1$ for different sample sizes and dependency levels. Grey box-plots in the background depict the distribution for the $\overline{\KL}$-optimal stopping iteration, instead of the one chosen by bootstrapping. Note that the y-axis limits change for the different dependency levels to enhance readability for the small values in the independent case.}
	\label{sim2_sdratio_cat} 
\end{figure} 

\begin{sidewaystable}
	\caption{Simulation overview}
	\label{table_simOverview}
	\vspace{.2cm}
	\footnotesize
	\begin{tabular}{|l|p{1.9cm}|r|r|p{1.3cm}|p{1.3cm}|r|p{1.0cm}|p{1.3cm}|p{1.0cm}|p{1.0cm}|p{1cm}|r|}
				& \textit{Model}  & $n_{eff}$ & $n_{cov}$ & $N$ & $G$ & \textit{Dependency} & $\df$ & $\nu$ & \textit{CV type} & \textit{CV folds} & \textit{Method} & $\bar{\sigma}_\mu^2$ \\
		\hline
		Default & continuous & 40 		& 5			& 100 & 100 & In, Dep, Hi	  & 13	  & 0.2    & boot	& 10	& noncyc.    & 1					    \\
		\hline\hline
			Fig. \ref{sim2_sdratio_cat}, \ref{RMSE_comparison}	& continuous, \mbox{categorical} &    		&  			& 50, 100, 334, 500 &  &   & 	  &     &  		&  	&  	    &     					  \\
		\hline	
		Fig. \ref{sim2_sdratio_cat}, \ref{RMSE_comparison}	&  	&    		&  			& 50, 100, 334, 500 &  &   & 	  &     &  		&  	 &  	   &  2   	  \\
		\hline	
		Fig. \ref{sim2_grid_size}	&	&    		&  			&  & 20, 50, 100, 150 &    & 	  &     &  		&  	 &  	   &     	    \\
		\hline
		Fig. \ref{sim2_cv}	&  	&    		&  			&  &  & In, Hi  & 	  &     & kfold, boot, subs	& 10, 25, 50 	  &  	  &     		  \\
		\hline				
		Fig. \ref{sim2_dependency_df_nu}	& 		&   6 		&  	3		 & 		&  	& In, Dep, Hi   & 	13, 15, 17  &  0.05, 0.1, 0.2, 0.3   &  &  	 		&  	   & 			   \\
		\hline	
		\begin{tabular}{l}
			Fig. \ref{sim2_kldcvpaths}, \ref{sim2_methods_computationtime} \\ Tab. \ref{tab_methods}
			\vspace*{-.5cm}
		\end{tabular}	&  		&   10 		&  	3		& 	 & 			&    & 	  &     &  		&  	   & noncyc, cyclic    &  	      \\
		\hline\hline
		Main man. Fig. 5	& continuous 	& 40 		& 5			& 100, 334 & 104  & In, Dep, Hi	  & --	  & --   & --		& --	&  	refund     & 	1				    \\
		\hline\hline
		Fig. \ref{appSim_RMSE}	& application 	& 1 		& 100			& 334 & 104  & In, Dep, Hi	  & 15	  & 0.1    & boot		& 10	&  	noncyc.     & 	--				    \\
		\hline						
	\end{tabular}
	
	\vspace{.5cm}
	Each row in the table corresponds to a simulation run, where for each of the $n_{eff}$ true effect sets and $n_{cov}$ covariate sets all combinations of the named simulation parameter specifications are compared. If nothing is specified explicitly, the default is applied. Besides the non-cyclic and cyclic boosting method one simulation run is also performed with penalized maximum likelihood, which is indicated by method 'refund'. \textit{CV type} and \textit{CV folds} describe the type of resampling method applied and the respective folds. 'kfold' denotes k-fold cross-validation, 'boot' bootstrapping, and 'subs' sub-sampling, each performed on curve level. In each fold of sub-sampling, the data is randomly devided into 50\% training and 50\% test data. $\bar{\sigma}_\sigma^2$ is always set to 1.
	
\end{sidewaystable}

\begin{figure}
	\includegraphics[width = \textwidth]{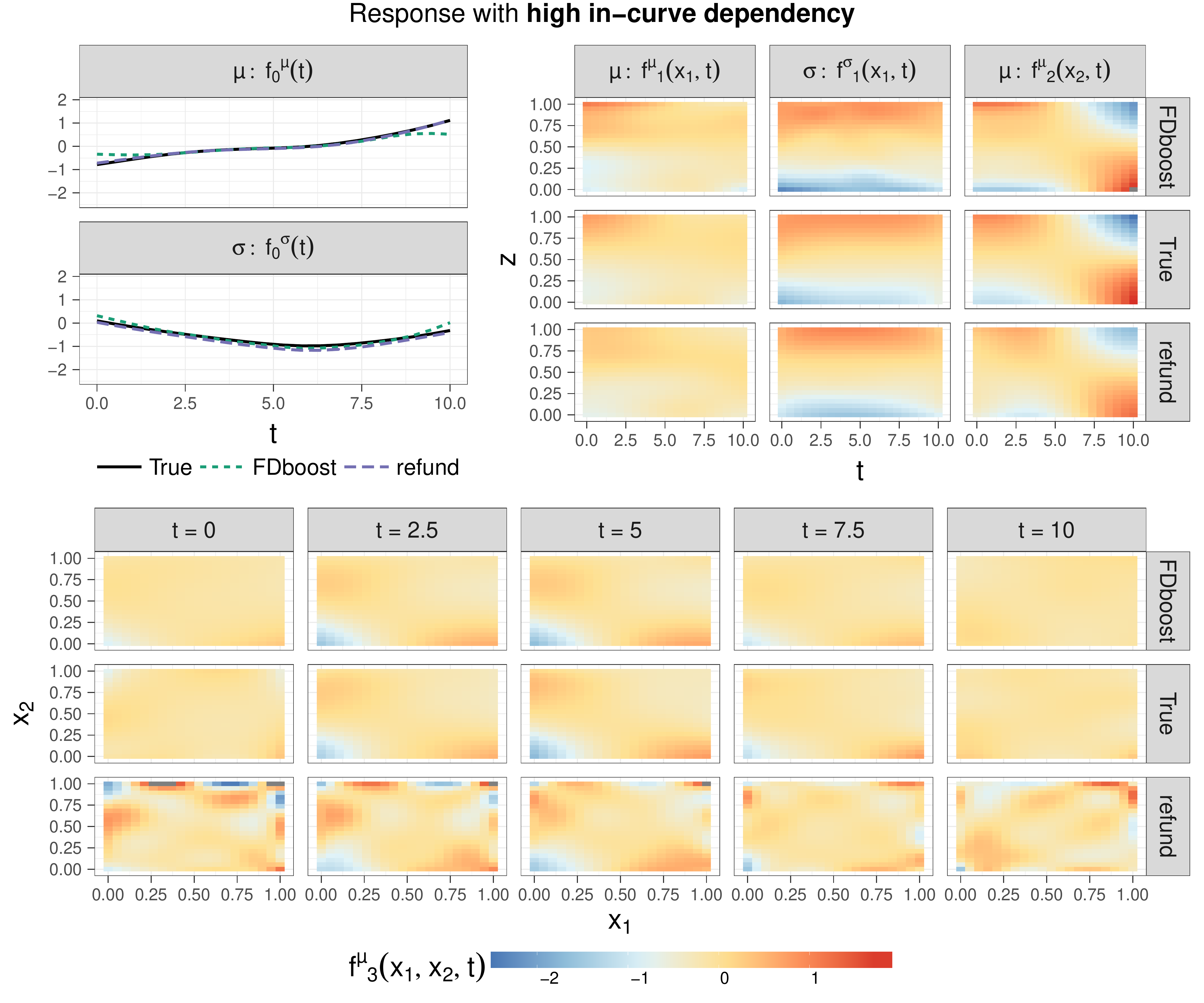}
	\caption{Example fit for high response in-curve dependency.
		The figure shows an example of true effect functions and estimates with the gradient boosting (\texttt{FDboost}) and penalized likelihood (\texttt{refund}) methods for the simulated continuous model scenario. The fit is based on a moderate number of $N = 334$ response-curves with $G = 100$ highly dependent measurements per curve and respective covariate samples. As the covariate interaction depends on $z_1,\ z_2$ and $t$, its effect functions are plotted for five fixed time points \textit{(bottom)}. Especially for this complex effect, we see how the regularization via curve-wise bootstrapping for \texttt{FDboost} prevents the over-fitting that might occur with \texttt{refund}, if the estimated effects are too complex in relation to the given sample size of curves.}
	\label{fig_fit_comparison_refund}
\end{figure}

\begin{figure}[H]
	\centering
	\includegraphics[page = 1, width = .75\textwidth]{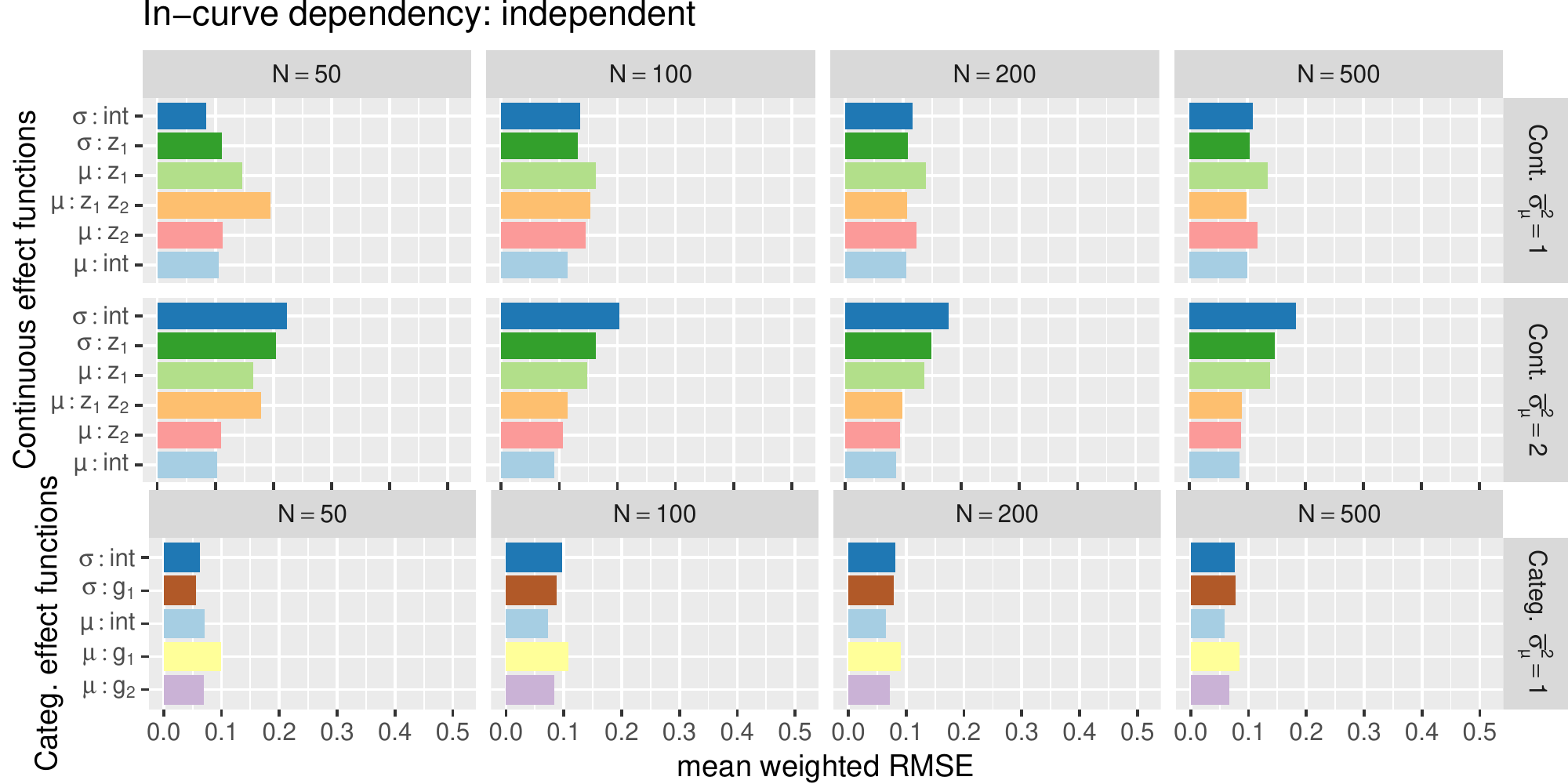}
	\includegraphics[page = 2, width = .75\textwidth]{Simulation2_weightedRMSE}
	\includegraphics[page = 3, width = .75\textwidth]{Simulation2_weightedRMSE}
	
	\caption{Weighted RMSE for individual effects in the continuous and categorical model:
	Comparison of goodness-of-fit for effect functions in the continuous models with $\bar{\sigma}_\mu^2 = 1, 2$ and for the categorical model with $\bar{\sigma}_\mu^2 = 1$ for different sample sizes and dependency levels. To account for their scaling, individual RMSE values are presented relative to the mean variance of simulated effect functions, i.e. relative to $\bar{\sigma}_\mu^2$ and $\tau(\bar{\sigma}_\sigma^2)$, respectively. For each effect function, the mean weighted RMSE over the simulations is depicted.}
	\label{RMSE_comparison}
\end{figure}

\begin{figure}[H]
	\includegraphics[width = \textwidth]{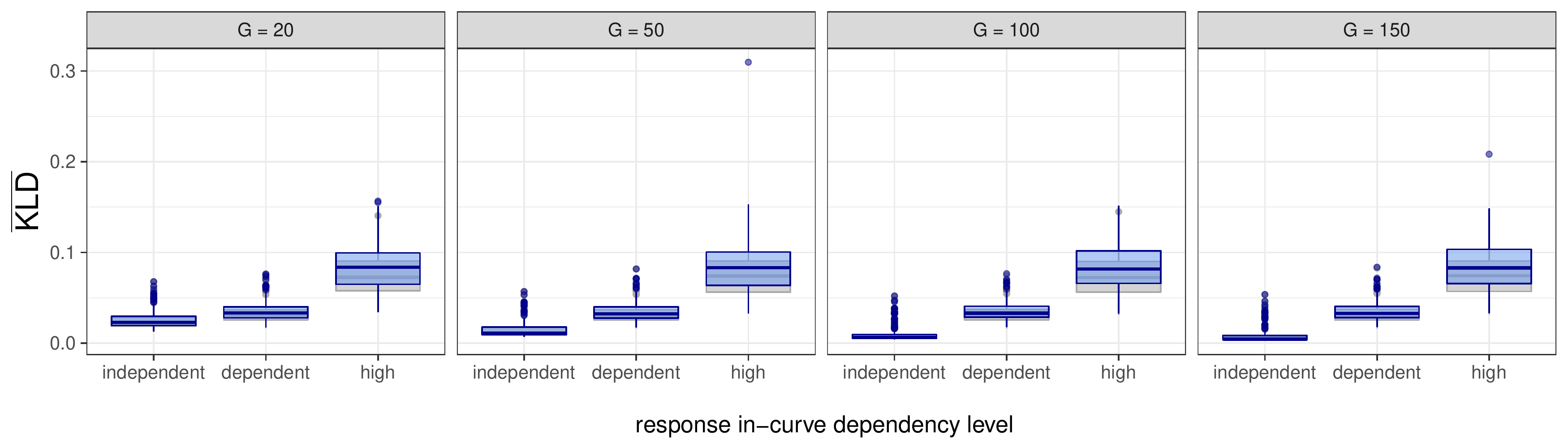}
	
	\caption{Comparison of different grid sizes G:
	Grid sizes G = 20, 50, 100, and 150 are compared for different dependency levels keeping $N=100$ fixed. Box-plots visualize the distribution of $\overline{\KL}$ for fitted models \textit{(blue, foreground)}. We only observe a considerable effect for independent response measurements. The $\overline{\KL}$ that would have been achieved with an optimal stopping iteration is depicted in the background \textit{(gray)}.}
	\label{sim2_grid_size}
\end{figure}

\begin{figure}[H]
	\centering
	\includegraphics[width=\textwidth]{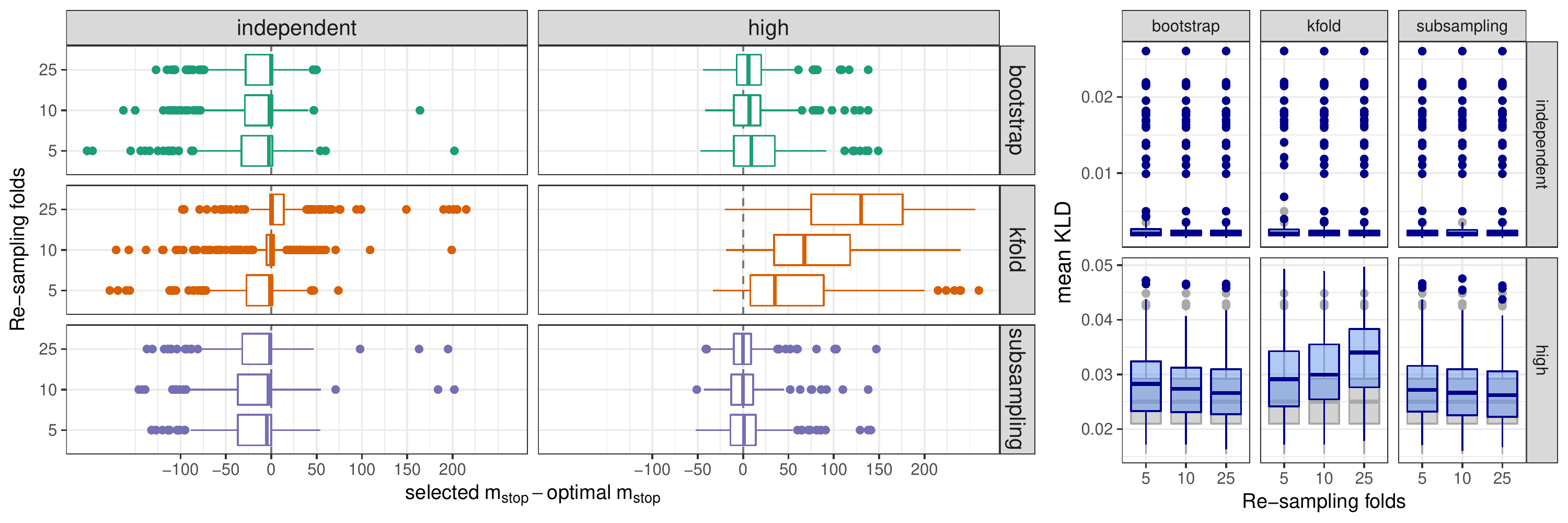}

	\caption{Comparison of re-sampling methods:
	Box-plots summarizing the deviation of selected stopping iterations $m_{stop}$ from the KLD-optimal $m_{stop}$ in the simulation. We compare bootstrapping, k-fold cross-validation, and sub-sampling for no and high in-curve dependency. Each method is applied with 5, 10, or 25 folds re-sampling. In sub-sampling, the data is randomly split into 50\% training set, 50\% test set in each fold. Dashed gray lines depict zero deviation. Since contribution of base learners decreases with boosting iterations, deviations are worse for high in-curve dependence.}
	\label{sim2_cv}
\end{figure}

\begin{figure}[H]
	\includegraphics[page = 1, width = .5\textwidth]{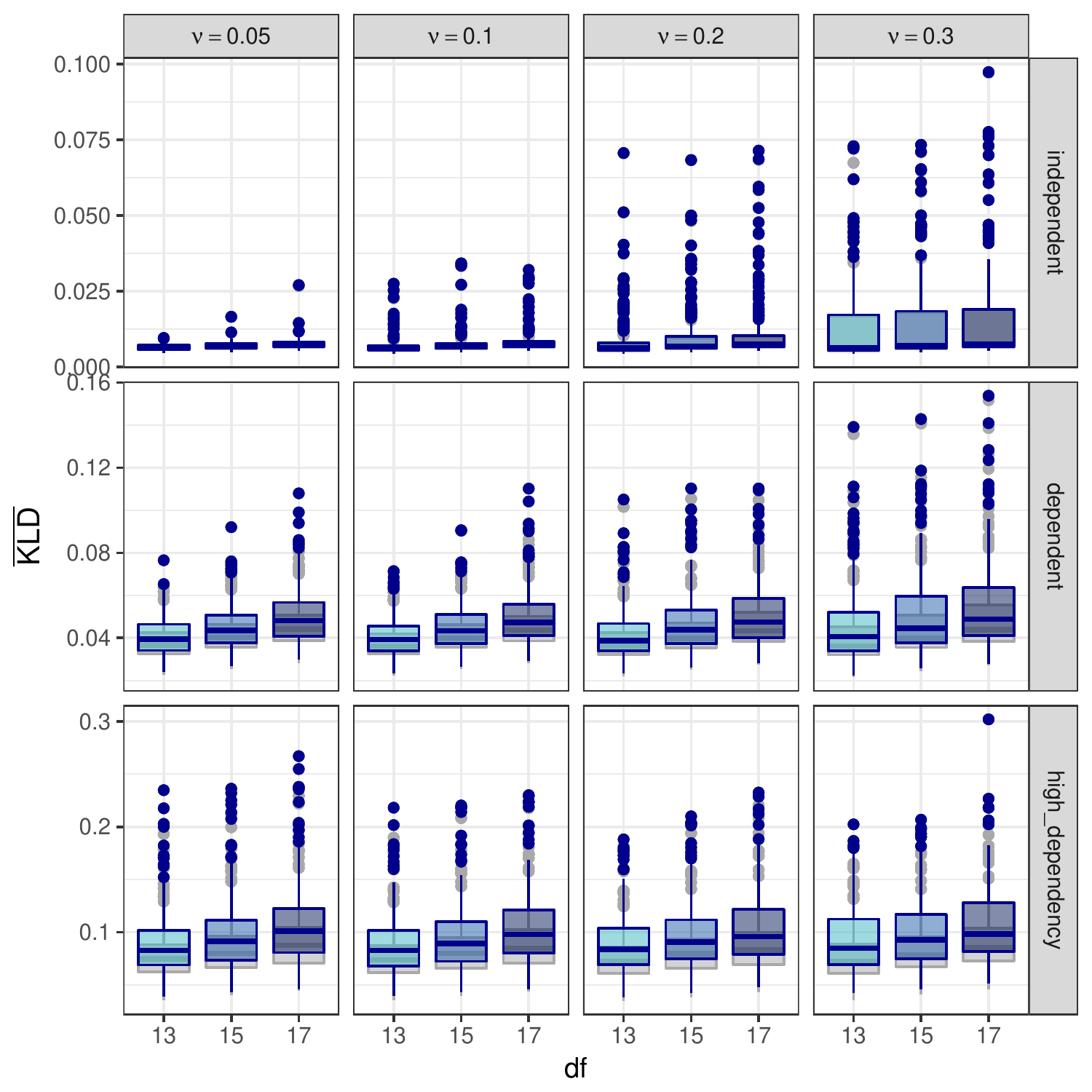}
	\includegraphics[page = 2, width = .5\textwidth]{Simulation2_dependency_df_nu}
	
	\caption{Comparison of different $\df$ and $\nu$ combinations:
	Base learner degrees of freedom $\df$ and step-length $\nu$ determine the flexibility of base learners in each boosting iteration. We compare different combinations for different dependency levels. To this end, distributions of $\overline{\KL}$ \textit{(left)} and corresponding stopping iterations $m_{stop}$ \textit{(right)} in simulations are displayed with box-plots. For the $\overline{\KL}$, grey boxes in the background correspond to the $\overline{\KL}$ for the optimal $m_{stop}$. We identify $\df=13$ and $\nu = 0.2$ to be a suitable and yet fast combination. We  do not consider $\df<13$, as our model effects are rather complex, and we do not want to apply $\df$ too close to the dimension of the penalty null space of the P-splines.  E.g., for the base-learner reflecting the smooth interaction between $z_1$ and $z_2$ over $t$, we distribute the degrees of freedom such that we obtain $\df=3$ for the P-spline in direction of $t$ (with penalty null space dimension $2$) and $\df=\sqrt{13/3}\approx 2.1$ in direction of $z_1$ and $z_2$, respectively.}
	\label{sim2_dependency_df_nu} 
\end{figure}

\begin{table}[H]
	\caption{Mean $\overline{\KL}$ for available gradient boosting methods for GAMLSS}
	\label{tab_methods}
	\centering
	\begin{tabular}{l|c|c|}
					& cyclic & non-cyclic \\	\hline
		independent & 0.1764 & 0.0067 \\ \hline
		dependent	& 0.2410 & 0.0353 \\ \hline
		high dependency & 0.3078 & 0.1010 \\ \hline
	\end{tabular}
	
	\vspace{.3cm}
	\footnotesize
	Mean $\overline{\KL}$ for available boosting methods and for different levels of in-curve dependency. Method 'cyclic' was first proposed by \citet{MayrSchmid2012}. Later, non-cyclic methods were developed by \citet{Thomas2016}.
	For all dependency levels, we observe that method 'non-cyclic' exhibits a better $\overline{\KL}$-performance than 'cyclic'. 
\end{table}


\begin{figure}[H]
	\centering
	\includegraphics[page = 1, width = .8\textwidth]{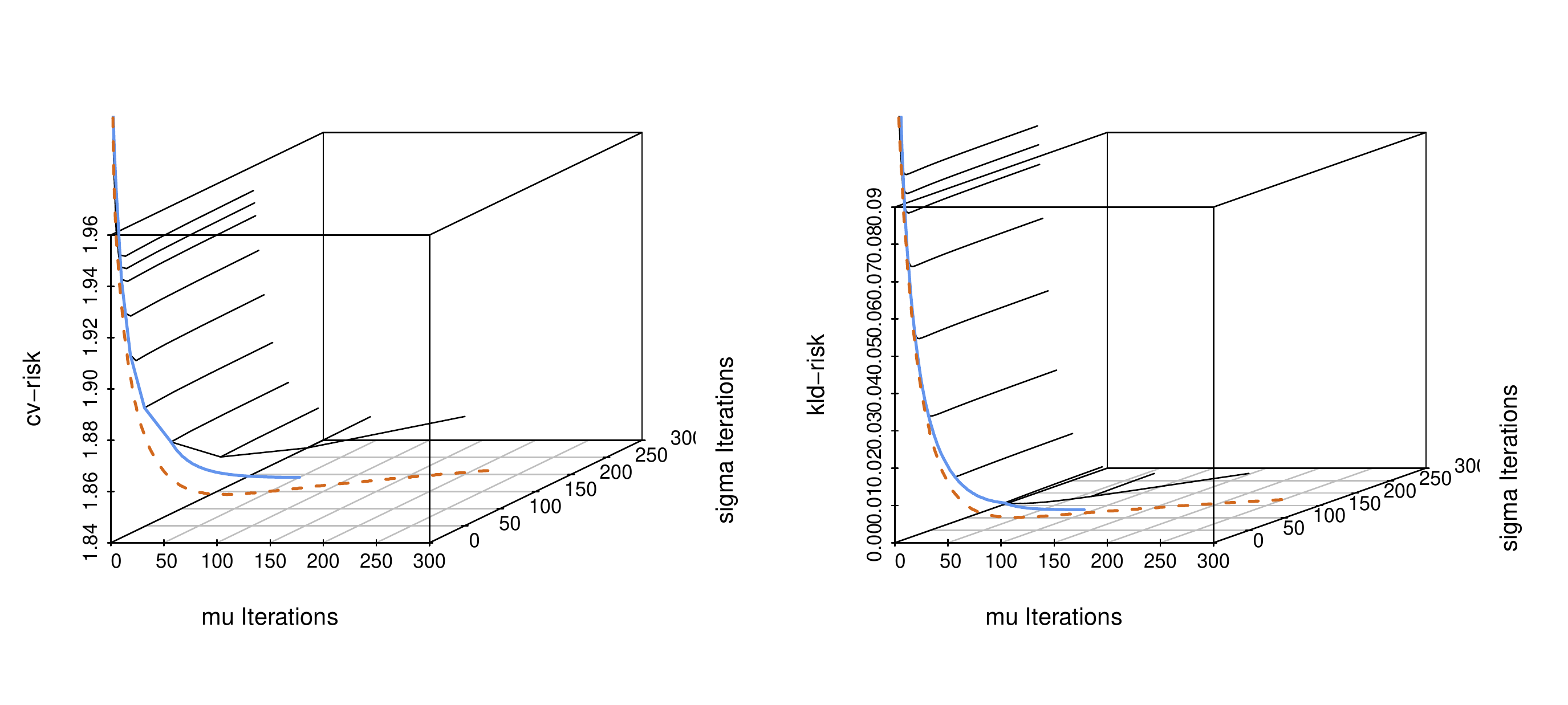}
	\includegraphics[page = 2, width = .8\textwidth]{Simulation2_kldcvpaths_sp3D}
	\includegraphics[page = 3, width = .8\textwidth]{Simulation2_kldcvpaths_sp3D}
	
	\caption{Cross-validation and $\overline{\KL}$ paths for the two different GAMLSS boosting methods:
	Exemplary cross-validation and $\overline{\KL}$ paths for both available gradient boosting methods for GAMLSS. x- and y-axes correspond to boosting steps in direction of $\mu$ and $\sigma$. The z-axis corresponds to resulting cross-validation error or $\overline{\KL}$, respectively. We compare cross-validation and $\overline{\KL}$ paths for in-curve dependency level \textit{independent} \textit{(top)}, for  \textit{dependent (middle)}, and for \textit{high dependency (bottom)}. In method 'cyclic', $\mu$- and $\sigma$-base learners are either updated alternately, or, from a certain point, the learners for one parameter are updated only \textit{(black; blue indicates selected path)}. In the 'noncyclic' method \textit{(chocolate, dashed)} free paths can be chosen. We observe, that the cross-validation error nicely approximates the structure of the $\overline{\KL}$. Especially for high in-curve dependencies, the $\overline{\KL}$ drops fast in the beginning, but rises afterwards when over-fitting occurs.}
	\label{sim2_kldcvpaths}
\end{figure}

\begin{figure}[H]
	\centering
	\includegraphics[width = .6\textwidth]{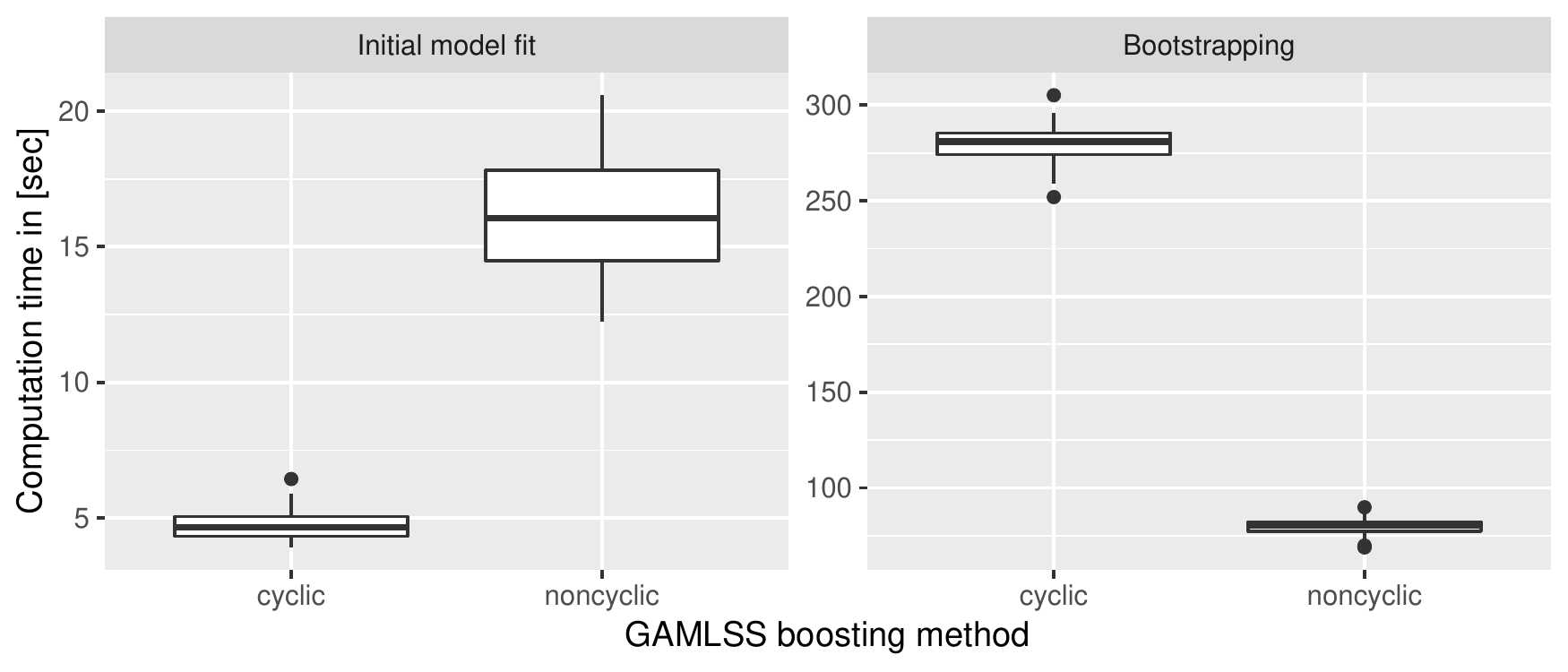}
	
	\caption{Computation time for available gradient boosting methods for GAMLSS:
	Computation time for fitting the continuous model with a maximum of $m_{stop} = 400$ boosting iterations. The cyclic and non-cyclic GAMLSS boosting methods are compared with respect to required time for a single model fit and for 10-fold bootstrapping. The models were fitted on a 64-bit linux server. While a single model fit in the cyclic method is faster, as the base-learners for the mean and standard deviation are fitted in an alternating way, re-sampling methods, like bootstrapping, take a lot longer than with the noncyclic method, since a seperate $m_{stop}$ for each parameter has to be chosen, which demands for multiple model estimations per re-sampling fold.}
	\label{sim2_methods_computationtime} 
\end{figure}

\begin{figure}[H]
	\includegraphics[page = 3, width = \textwidth]{appSim_rMSE}
	\includegraphics[page = 1, width = .5\textwidth]{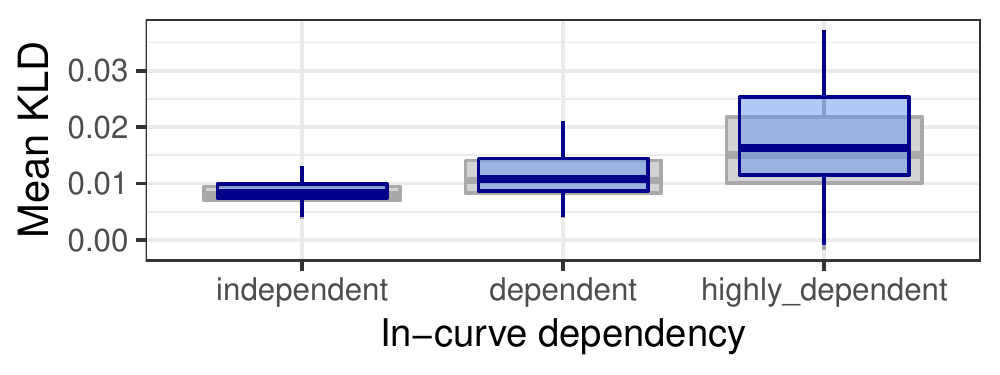}
	\includegraphics[page = 2, width = .5\textwidth]{appSim_diagnostics}
		
	\caption{Diagnostic plots for the application-motivated simulation:
	\textit{Top:} for each covariate effect in the model and each in-curve dependency level, we consider the mean RMSE over 120 simulation runs. Effect functions are subdivided into those for the mean $\mu$, those for the standard deviation $\sigma$, those for the zero-area probability $p$ and those for the scale parameter $\nicefrac{\sigma}{\mu}$. For $p$ both a a step-function intercept (StepI) and smooth intercept (SmthI) centered around StepI are included into the model formula. However, SmthI was never selected in the original model fit. \textit{Bottom, left:} box-plots for $\overline{\KL}$ \textit{(blue, foreground)} and the optimal $\overline{\KL}$ \textit{(gray, background)}. \textit{Bottom, right:} box-plots for stopping iteration $m_{stop}$ determined by bootstrapping \textit{(blue, foreground)} and the $\overline{\KL}$-optimal $m_{stop}$ \textit{(gray, background)}. While in this setting the optimal stopping iteration is typically overestimated by the bootstrap, this has only a slight effect on the fitting quality, as the $\overline{\KL}$ achieved by the estimation is very close to the optimal. This reflects the $\overline{\KL}$ paths in dependence of $m_{stop}$ having a very flat minimum. }
	\label{appSim_RMSE}
\end{figure}

\section{Application details}
\label{app_app}

\subsection{Data}
\label{data_preparation}

\begin{figure}[H]
	\includegraphics[width = .541\textwidth]{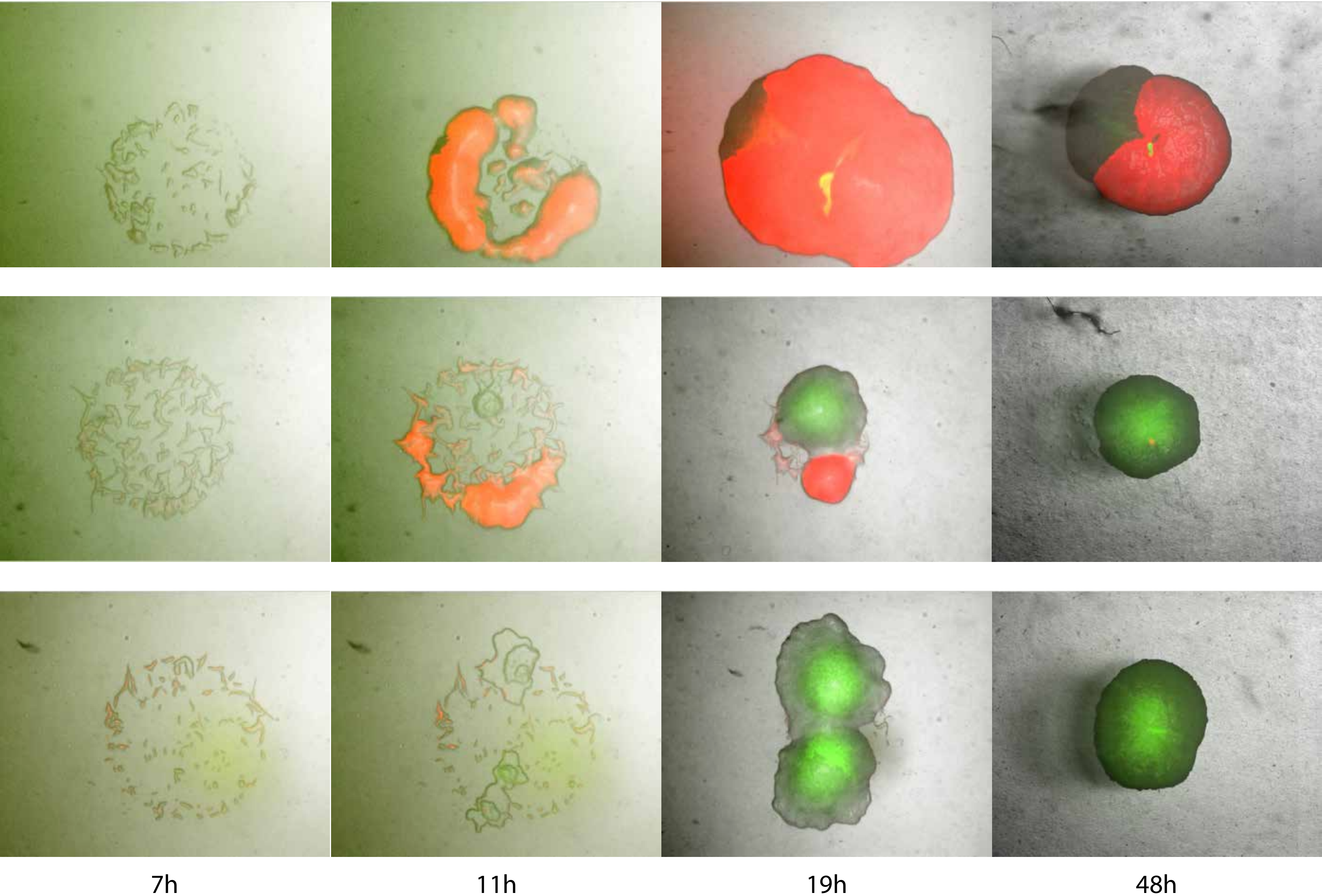}
	\includegraphics[width = .459\textwidth]{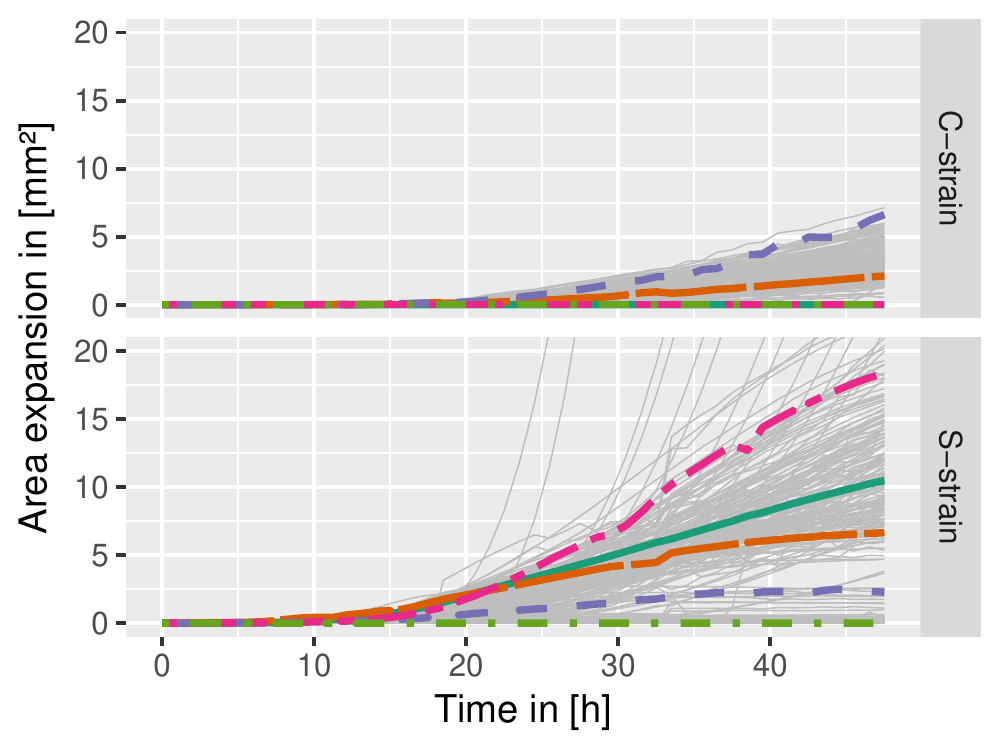}
	
	\caption{Bacterial growth data:
	\textit{Left:} Overlay of bright-field, red and green fluorescence for three different bacterial spots after 7, 11,  19, and 48 hours. For the last two observation times the microscope zoom level was adjusted to cover the full bacterial area. \textit{Right:} Bacterial growth curves of the C-strain \textit{(top)} and S-strain \textit{(bottom)}. Highlighted C- and S- example curves from the same spot are marked correspondingly.}
	\label{fig_growthCurves}
\end{figure}

The data set contains $N=334$ observed functional response curves $S_i(t)$ and covariate curves $C_i(t)$. Their value corresponds to area expansion in $mm^2$ of S- and C- bacterial strain, respectively. Original values in $\mu m^2$ were converted to avoid numerical instability. The curves are measured on a common grid of length $G=105$ in the time interval $T=[0,48\, h]$. For the first $18\, h\ 30\, min$, 75 measurements are taken every $15\, min$. The remaining 29 measurements are taken every hour. S- and C- strain are distinguished via red and green fluorescence. \citet{vonBronk2016} employ automatic segmentation of propagation areas from different color channels of recorded microscope pictures. In order to capture the full area over the whole time span, four different microscope zoom levels are applied. After $12\, h\ 15\, min$ the zoom level of the microscope is adjusted for the first time, again after $18\, h\ 30\, min$ and after $33\, h\ 30\, min$. The MitC concentration is included as a categorical covariate with four levels. As the covariate $C'(t)$ is calculated from $C(t)$ by numerical differentiation, the last time point at $t=48\,h$ is dropped for all other curves. For every observation a positive amount of $S$- and $C$-cells is present at the beginning. Due to automatic area segmentation some growth curves contain outliers marked by distinct jumps in the growth. Corresponding values of the cures were identified manually, deleted and replaced by spline interpolation. For each MitC concentration, two experimental batches were conducted. For MitC = 0 these include 34 and 41 bacterial growth spots; for MitC = 0.005 47 and 40 spots; for MitC = 0.01 45 and 40 spots; and for MitC = 0.1 46 and 41 spots. The number of spots per experimental batch varies as only spots with a positive number of S- and C-cells in the beginning were chosen. Figure \ref{fig_growthCurves} shows example microscope pictures of bacterial colonies and an overview over S- and C-growth curves.


\subsection{Comparison with usual growth models}
\label{app_modelcomparison}

GAM(LSS) provide flexible means of modeling growth curves. We compare them to four parametric growths models. The four parameter Baranyi-Roberts model \citep{Baranyi1994} and the Gompertz model \citep{Gibson1988} in the parametrization of \citet{Zwietering1990} present two popular approaches to modeling bacterial growth. \citet{Baty2014} formulate these in terms of common parameters $y_0, y_\infty, \mu_{max}$ and $L$. The Baranyi-Roberts model is given by
\begin{equation*}
y(t) = y_0 + \log_{10}\left(\frac{-1 + \exp(\mu_{max} L) + exp(\mu_{max} t)}{exp(\mu_{max} t)} - 1 + 10^{y_\infty - y_0} exp(\mu_{max} L)  \right).
\end{equation*}
The Gompertz model is given by
\begin{equation*}
y(t) = y_0 + \left(y_\infty - y_0\right) \exp\left(-\exp\left( \frac{\mu_{max} e(L - t)}{(y_\infty - y_0)\log(10)} + 1\right) \right).
\end{equation*}

\citet{Weber2014} employ a five-parameter sigmoidal function describing effective radii of bacterial colonies in bacterial interaction. Depending on parameters $a,\dots, d$ and $y_0$, their model takes the form
\begin{equation*}
y(t) = a + \frac{y_0 - a}{\left(1+(t/c)^b\right)^d}
\end{equation*} 

We also add the standard three parameter logistic growth model of the form 
\begin{equation*}
	y(t) = \frac{y_\infty y_0\exp(rt)}{y_\infty + y_0 \left(\exp(rt)-1\right)} .
\end{equation*}
 
\begin{figure}[H]
	\centering
	\includegraphics[width = .77\textwidth]{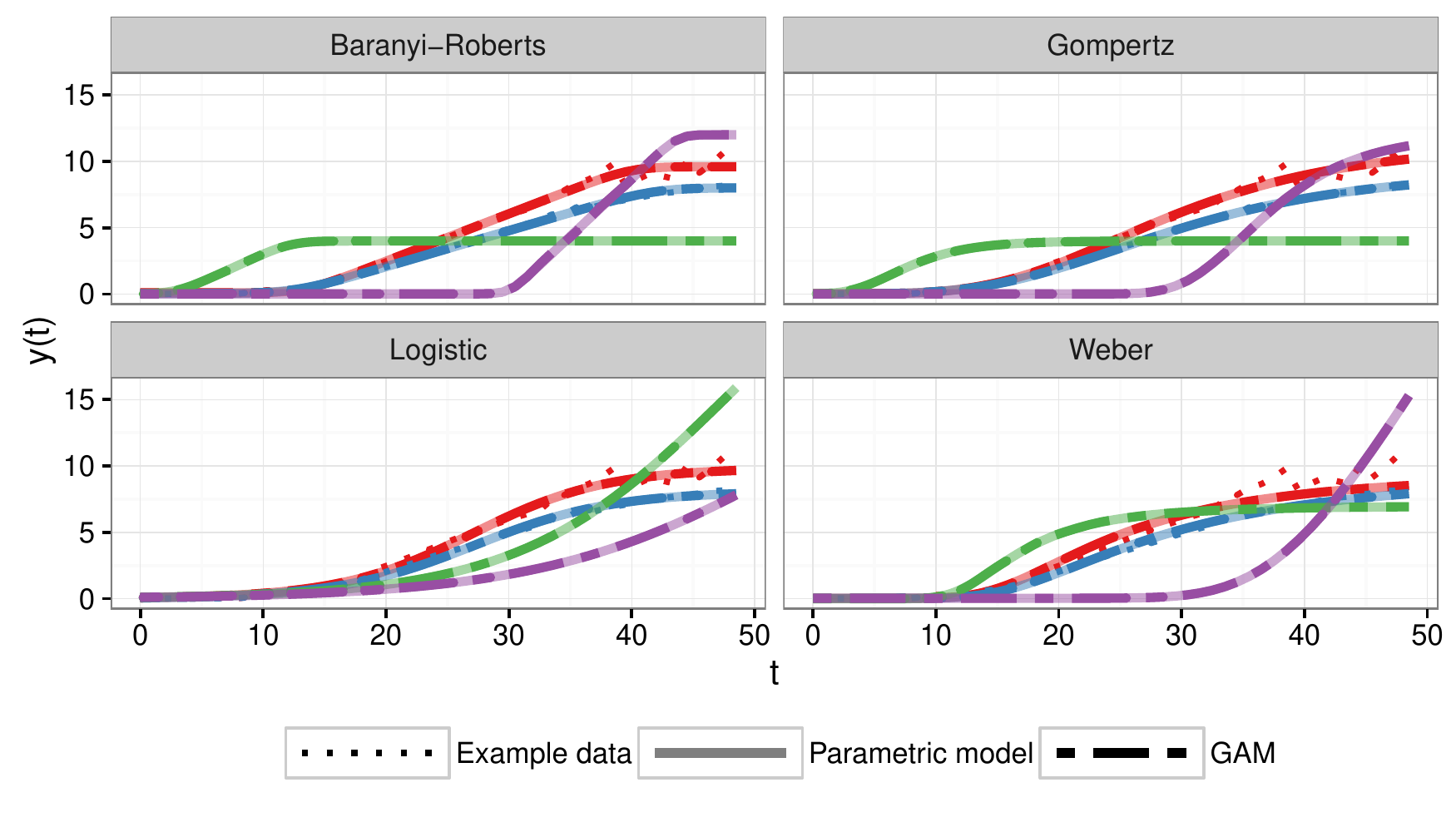}
	
	\caption{Comparison of growth models:
	GAM approximations to different curves of parametric growth models. For the red and blue curves the parametric models were fitted to example data. Green and violet curves display alternative growth curve shapes.}
	\label{fig_compare}
\end{figure}

We fit two example S-strain growth curves from our data set with each of the models in order to obtain realistic parameter values. Fitting is done via least squares. In addition, we pick two alternative parameter settings. For comparison, we fit each of the generated parametric model curves with GAM. In analogy to our applied model, we fit the GAM by boosting using a gamma distribution loss with a log-link and a functional intercept constructed with a B-spline. Resulting GAM approximation turns out to be nearly perfect for all parametric models (Fig. \ref{fig_compare}).

\subsection{Estimated effect functions and bootstrap uncertainty bounds}
\label{app_predictor}

This section contains supplementary figures showing all estimated effect functions on predictor level (Figure \ref{fig_intMitC} and \ref{Ceffects}). In addition we computed point-wise bootstrap uncertainty bounds for all  effects in a basic bootstrap 95\% confidence interval type procedure with 1066 bootstrap samples. Accordingly, the intervals are computed as $[\hat{f}\jq(t) - \Delta^{*}_{0.975}(t), \hat{f}\jq(t) - \Delta^{*}_{0.025}(t)]$ for group-specific functional effects and $[\hat{\beta}\jq(s,t) - \Delta^{*}_{0.975}(s,t), \hat{\beta}\jq(s,t) - \Delta^{*}_{0.025}(s,t)]$ for historical effect coefficient surfaces, where $\Delta^{*}_\alpha$ denotes the bootstrap estimate of the point-wise $\alpha$-quantiles of the distribution of $\hat{f}\jq-f\jq$ or $\hat{\beta}\jq-\beta\jq$, respectively.  The bounds are meant to give indications on the estimation precision complementing the results of our simulation study, but due to the complexity of the matter we refrain from interpreting them as valid confidence bounds. As the estimators are subject to shrinkage bias introduced by early stopping of the boosting algorithm, we can not expect the bootstrap estimators $\Delta^{*}_\alpha$ to be unbiased. In addition, we are limited to basic bootstrap confidence intervals as we do not have proper estimates for the estimators' variances and, thus, cannot compute studentized bootstrap or accelerated bias-corrected bootstrap intervals (compare, e.g.,  \citep{Hall1988}) without tremendous computational burden.  Visualizations of the uncertainty bounds can be found in Figure {\ref{fig_intMitC} for $f\jq$ and in Figures \ref{fig_3Dhistorical}, \ref{fig_historical} and \ref{fig_historical2} for $\beta\jq(s,t)$.\newline
Figure \ref{fig_overall_MitC} illustrates the effects of MitC on the overall mean and standard deviation, rather than the effect on the conditional mean and standard deviation given $S(t)>0$.

	
	


\begin{figure}[H]
    \centering
    \includegraphics[height = .9\textheight]{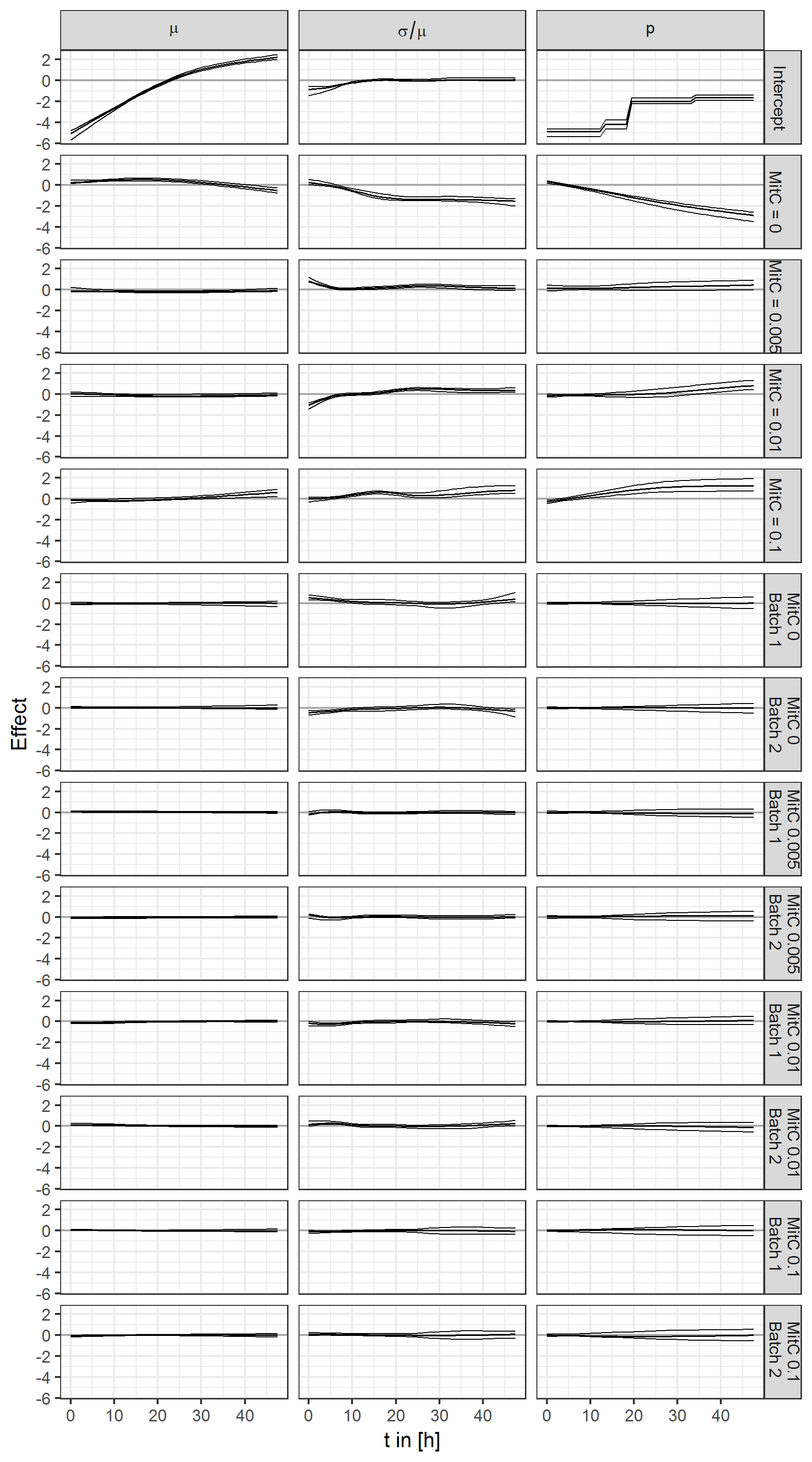}
    \caption{Estimated effect functions of the functional intercept, the MitC effects and the batch effects on $\mu$, $\nicefrac{\sigma}{\mu}$ and $p$ with 95\% bootstrap confidence interval type uncertainty bounds based on 1066 bootstrap samples.}
    \label{fig_intMitC}
\end{figure}

\begin{figure}[H]
	\centering
	\includegraphics[page = 5, width = .8\textwidth]{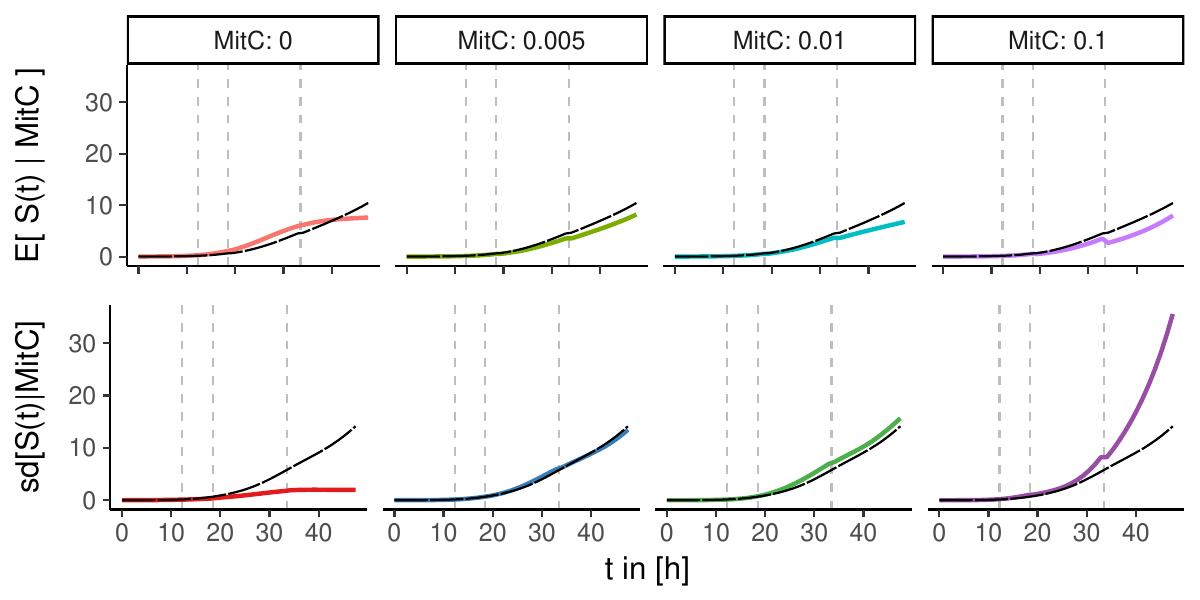}
	
	\caption{Overall MitC-effects: Estimated point-wise mean \textit{(top)} and standard deviation \textit{(bottom)} of the S-strain growth curves $S_i(t)$ without conditioning on $S_i(t)>0$. Long-dashed black curves correspond to the functional intercept, dashed vertical lines to the zoom level change-points.}
	\label{fig_overall_MitC}
\end{figure}


\begin{figure}[H]
	\includegraphics[page = 1, width = .45\textwidth]{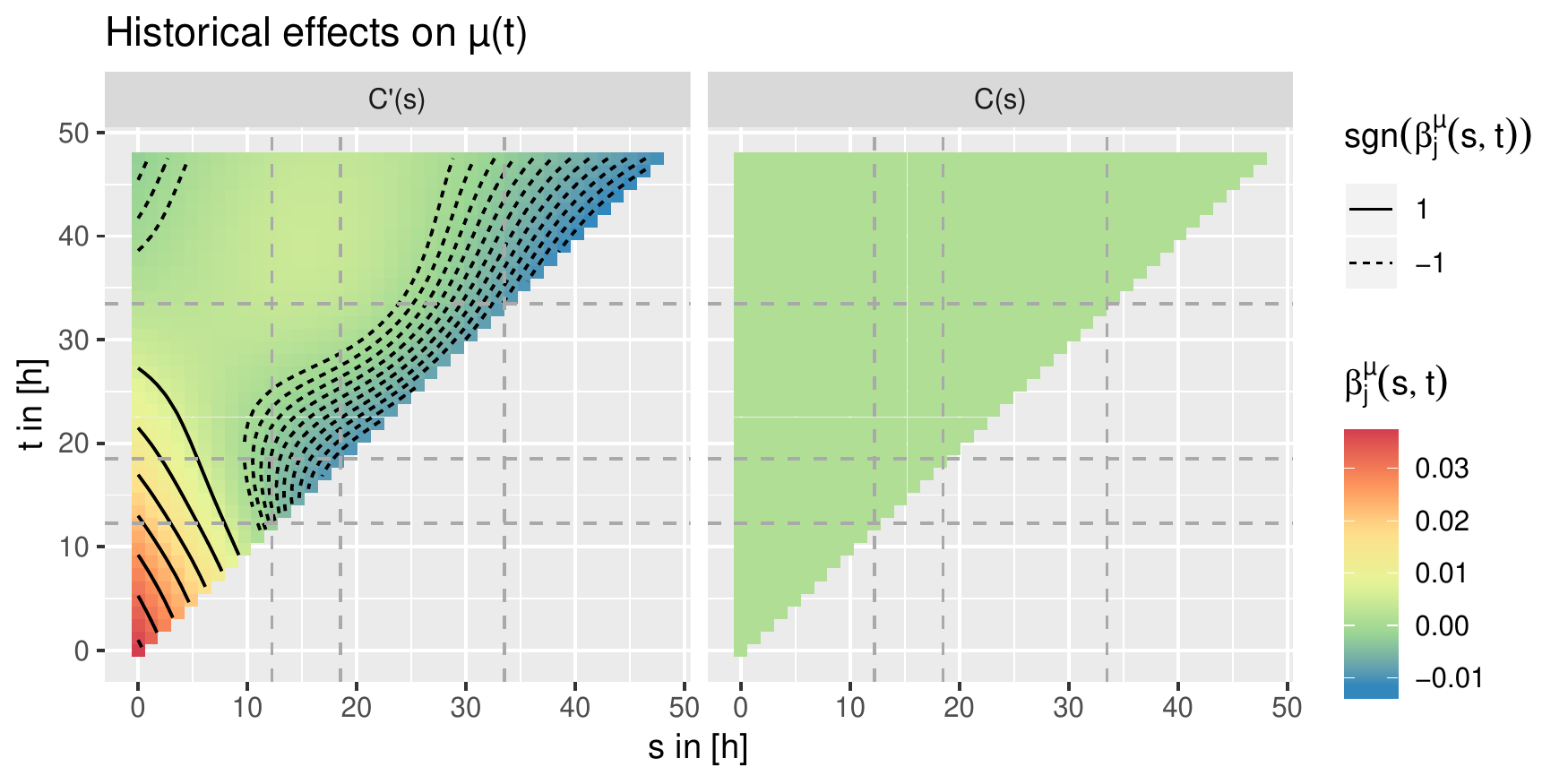}
	\includegraphics[page = 2, width = .45\textwidth]{C_effects}
	
	\includegraphics[page = 3, width = .45\textwidth]{C_effects}
	\includegraphics[page = 4, width = .45\textwidth]{C_effects}
	
	\caption{Historical effects of C-strain propagation area:
	Coefficient functions $\beta^{(q)}(s,t)$ for the historical effects of $C'(t)$ and $C(t)$ on the mean \textit{(top, left)}, the scale parameter \textit{(top, right)}, the zero area probability \textit{(bottom, left)}, and the standard deviation \textit{(bottom, right)} of the S-strain growth curves. The y-axis presents the time line for the S-strain curve, the x-axis the one for the C-strain. Grey dashed lines mark zoom breaks. Note that for in the early phase with $s,t\leq10\; h$ there are almost no observations with $S_i(t)=0$, so the corresponding effects on $p(t)$ should not be interpreted. Moreover, the $C-\mu$-effect was never selected by the boosting algorithm and is, thus, constantly zero.}
	\label{Ceffects}
\end{figure}

\begin{figure}[H]
    \centering
    \includegraphics[height = .3\textheight]{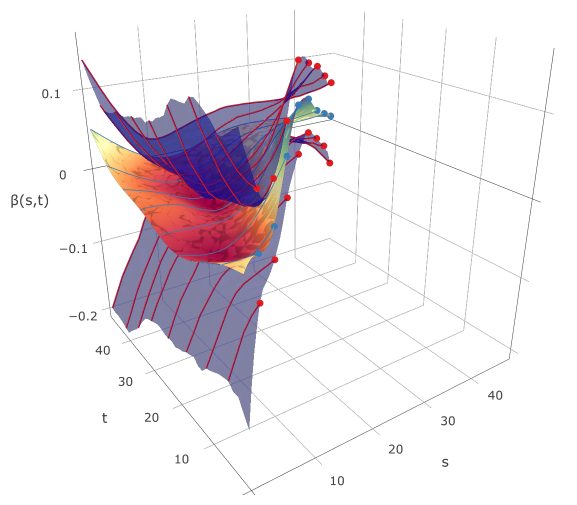}
    \caption{Example surface plot of $C'$-$\sigma$-effect with 95\% bootstrap confidence interval type uncertainty bounds visualizing the line segments in the Figures \ref{fig_historical} and \ref{fig_historical2}.}
    \label{fig_3Dhistorical}
\end{figure}

\begin{figure}[H]
    \centering
    \includegraphics[width = .8\textwidth]{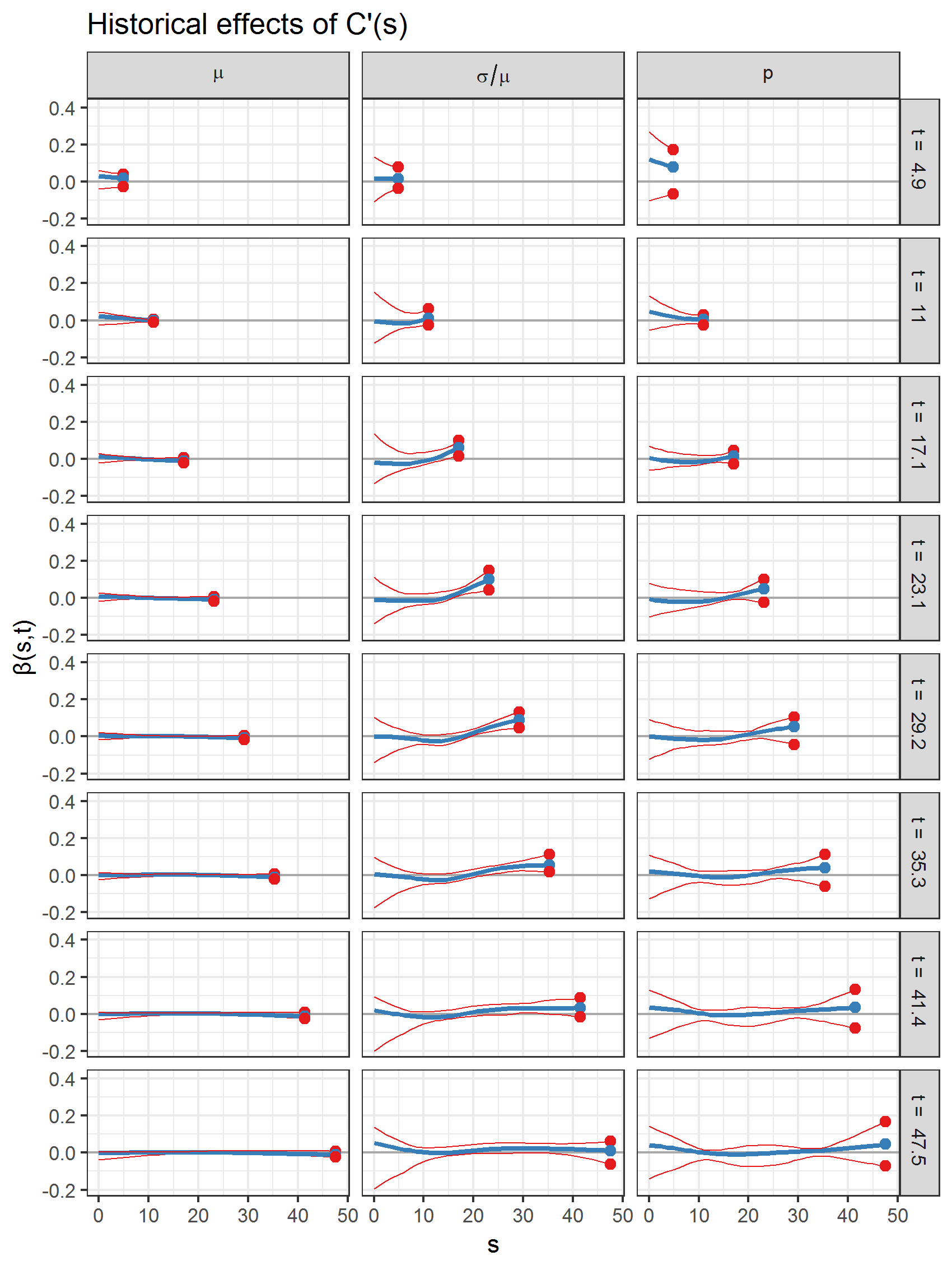}
    \caption{95\% bootstrap confidence interval type uncertainty bounds (\textit{red}) for estimated line segments $s \mapsto \hat{\beta}(s, t)$ of the historical effect of $C'(s)$ for different fixed values of $t$ (\textit{blue}). The dots correspond to the values at $s=t$ and therefore serve for identification (compare Figure \ref{fig_3Dhistorical}).}
    \label{fig_historical}
\end{figure}

\begin{figure}[H]
    \centering
    \includegraphics[width = .8\textwidth]{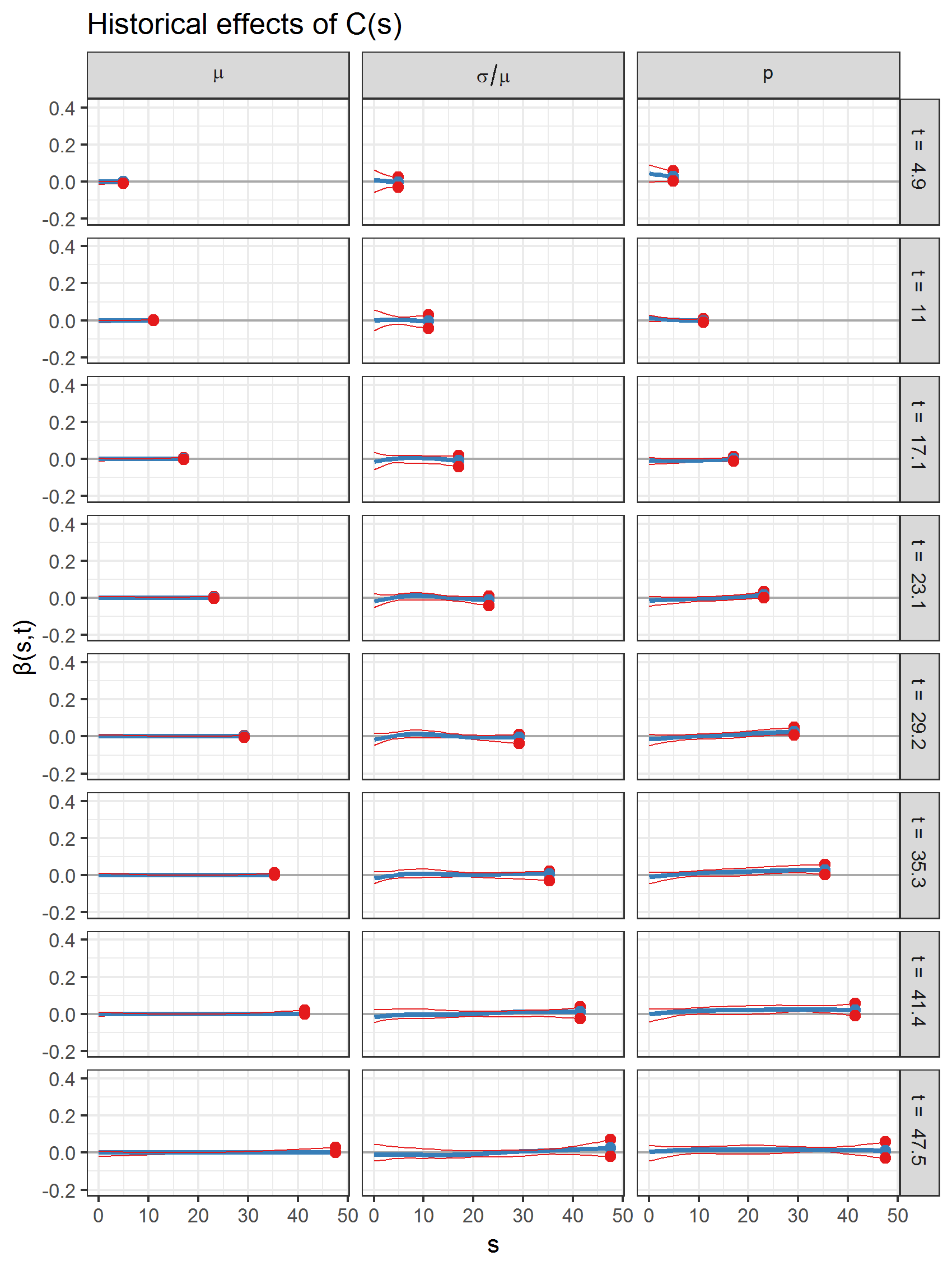}
    \caption{95\% bootstrap confidence interval type uncertainty bounds (\textit{red}) for estimated line segments $s \mapsto \hat{\beta}(s, t)$ of the historical effect of $C(s)$ for different fixed values of $t$ (\textit{blue}). The dots correspond to the values at $s=t$ and therefore serve for identification (compare Figure \ref{fig_3Dhistorical}).}
    \label{fig_historical2}
\end{figure}

%
%
%
%


\end{document}